\documentclass[11pt]{article}

\usepackage{amssymb,amsmath,amsfonts,amsthm}

 \usepackage[numbers,sort&compress]{natbib}
\usepackage{graphicx}

\usepackage{hyperref}
\hypersetup{
   colorlinks   =  true,
    linkcolor    = blue,
    citecolor    = red,
     urlcolor	=blue,     
 }

\usepackage{float}
\usepackage{placeins}

\input xy
\xyoption{all} 

\renewcommand{\theequation}{\arabic{section}.\arabic{equation}}

\newcommand\be{\begin{equation}}
\newcommand\ee{\end{equation}}
\newcommand\bea{\begin{eqnarray}}
\newcommand\eea{\end{eqnarray}}

\theoremstyle{plain}
\newtheorem{theorem}{Theorem}

\newtheorem{proposition}[theorem]{Proposition}

\theoremstyle{definition}

\numberwithin{theorem}{section}
\numberwithin{equation}{section}

\def\dd{{\rm d}}
\def\eee{{\rm e}}
\def\>#1{{\bf #1}} 

 \def\gam{\gamma}
 \def\xii{\Xi}
 \def\wcan{{\omega_{\rm can}}}
\def\aaa{\alpha}
 \def\rr{r}

\begin{document}

\
  \vskip0.5cm

 \noindent
 {\Large \bf  
{Generalized Buchdahl   equations as Lie--Hamilton systems  from \\[6pt] the `book' and oscillator algebras:} Quantum deformations and their\\[6pt]     general solution}

\medskip 
\medskip

\begin{center}

{\sc  Rutwig Campoamor-Stursberg$^{\,1,2}$, Eduardo Fern\'andez-Saiz$^{\,3}$\\[2pt] and Francisco J.~Herranz$^{\,4}$}

\end{center}

\medskip

 \noindent
$^1$ Instituto de Matem\'atica Interdisciplinar, Universidad Complutense de Madrid, E-28040 Madrid,  Spain

\noindent 
$^2$ Departamento de \'Algebra, Geometr\'{\i}a y Topolog\'{\i}a,  Facultad de Ciencias 
Matem\'aticas, Universidad Complutense de Madrid, Plaza de Ciencias 3, E-28040 Madrid, Spain

\noindent 
{$^3$ Department of Mathematics, CUNEF Universidad, 
C. de Leonardo Prieto Castro 2, 28040  Madrid, Spain}

\noindent
{$^4$ Departamento de F\'isica, Universidad de Burgos, 
E-09001 Burgos, Spain}

 \medskip
 
\noindent  E-mail: {\small
 \href{mailto:rutwig@ucm.es}{rutwig@ucm.es}, \href{mailto:e.fernandezsaiz@cunef.edu}{e.fernandezsaiz@cunef.edu}, \href{mailto:fjherranz@ubu.es}{fjherranz@ubu.es} 
}

\medskip

\begin{abstract}

\noindent
We revisit the nonlinear second-order differential equations 
$$
 \ddot{x}(t)=a (x )\dot{x}(t)^2+b(t)\dot{x}(t)
 $$
where $a(x)$ and $b(t)$ are arbitrary functions on their argument from the perspective of Lie--Hamilton systems. For the particular choice $a(x)=3/x$ and $b(t)=1/t$, these equations reduce to the Buchdahl equation considered in the context of General Relativity. It is shown that these equations are associated to the `book'  Lie algebra $\mathfrak{b}_2$, determining a Lie--Hamilton system for which the corresponding $t$-dependent Hamiltonian and the general solution of the equations are given. The procedure is illustrated considering several particular cases. We also make use of the quantum deformation of $\mathfrak{b}_2$ with quantum deformation parameter $z$ (where $q=\eee^z$), leading to a deformed generalized Buchdahl equation.  Applying the formalism of Poisson--Hopf deformations of Lie--Hamilton systems, we derive the corresponding deformed $t$-dependent Hamiltonian, as well as its general solution.  The generalized Buchdahl equation is further extended to the oscillator Lie--Hamilton algebra $\mathfrak{h}_4\supset \mathfrak{b}_2$, together with its quantum deformation, and the corresponding (deformed) equations are also analyzed for their exact solutions.  The presence of the quantum deformation parameter $z$ is interpreted as the introduction of an integrable perturbation of  the (initial) generalized Buchdahl equation, which is described in detail in its linear approximation.   Finally, it is also shown that, under quantum deformations, the higher-dimensional deformed generalized Buchdahl equations from either the book or the oscillator algebras do not reduce to  a sum of copies of the initial system but to intrinsic coupled systems governed by $z$.

  \end{abstract}
\medskip
\medskip


\medskip


\medskip

 \newpage


\tableofcontents

\newpage

 \section{Introduction}
\label{s1}

In a seminal paper, H.A.~Buchdahl  considered in 1964~\cite{Buchdahl} a class of metrics on a (3+1)-dimensional spherically-symmetric static spacetime in a General Relativity framework \cite{STE}. In terms of the so-called isotropic coordinates $(t,r,\theta,\phi)$,  this family of metrics was given  by
\be
\dd s^2=\frac{\bigl(1-f(r)\bigr)^2}{\bigl(1+f(r)\bigr)^2} \,\dd t^2 - (1+f(r))^4\left[\dd r^2 +r^2 \bigl(\dd\theta^2 +\sin^2\theta \dd \phi^2 \bigr) \right] ,
\label{aa0}
\ee
  where the function $f(r)$ had to be determined and units are chosen such that the   speed of light is normalized to $c=1$.  This equation was shown to correspond to a relativistic fluid sphere.
Under some additional requirements, the field equations gave rise to the following scalar second-order nonlinear ordinary differential equation (ODE in short) 
 \begin{equation}
 \frac{\dd^2 f}{\dd r^2}-\frac 3 f
 \left(\frac{\dd f}{\dd r}\right)^2-\frac 1r \,\frac{\dd f}{\dd r} =0,
 \label{aa1}
\end{equation}
which admits the exact solution (see \cite{Buchdahl}) 
\begin{equation}
f(r)=\frac{\pm 1}{ k_1\sqrt{1+k_2 r^2} } \, ,
\label{a2}
\end{equation}
where $k_1$, $k_2$ are the two integration constants. Note that equation (\ref{aa1})  can be written alternatively in terms of total derivatives as
\be
\frac{\dd }{\dd r}\left(\frac 1r\, \frac{\dd }{\dd r}\biggl( \frac1{f^2} \biggr) \right)=0 .
\label{a3}
\nonumber
\ee
The Buchdahl equation (\ref{aa0}) has also been considered in a  different context, namely that  of   nonlinear ODEs with time and position dependent coefficients, by replacing the radial function $f(r)$ by $x(t)$ and interpreting $t$ as the time~\cite{Duarte, Dua2, Chandrasekar, Nikiciuk}, leading to the modified Buchdahl equation 
  \begin{equation}
 \frac{\dd^2 x}{\dd t^2}=\frac 3 x
 \left(\frac{\dd x}{\dd t}\right)^2+\frac 1t \,\frac{\dd x}{\dd t}\, .
 \label{a1}
\end{equation}
In this form, the equation has been studied in the aforementioned works using several different methods: within {\em Example 2} in~\cite{Duarte}, as  {\em Example 1} in~\cite{Chandrasekar}, and in {\em Example 3.6} in~\cite{Nikiciuk}. 

In the general context of nonlinear ODEs, the equation \eqref{a1} is nothing but a special case of the generic equation 
\be
 \frac{\dd^2 x}{\dd t^2}=a(x)\biggl(\frac{\dd x}{\dd t}\biggr)^2+b(t)\,\frac{\dd x}{\dd t}\, ,
 \label{b01}
\ee
 where $a(x)$ and $b(t)$ are arbitrary functions of their arguments. This generalization can also be regarded as the equation of the motion of a   dynamical system with variable coefficients $a(x) $ and $b(t)$, depending on velocity/momentum terms. It is not difficult to see that, dividing by $\dot{x}=\displaystyle\frac{\dd x}{\dd t}$,\footnote{In other terms, the equation admits an integrating factor.} the equation can be rewritten as 
\be
 \frac{\dd }{\dd t} \ln |\dot{x}|  =\frac{\dd }{\dd t}\left(\int^{x} a(\xi) \dd \xi\right) +\frac{\dd }{\dd t}\left( \int^{t} b(\tau)\dd \tau \right),
 \label{b02}
\ee 
showing that \eqref{b01} is an exact equation, the solution of which is recovered from the expression
\be
\ln |\dot{x}|= \int^{x} a(\xi) \dd \xi  +\int^{t} b(\tau)\dd \tau +k_1
 \label{b03}
\ee  
using quadratures, where $k_1$ is a constant of integration \cite{Ince,KAM,C132}. It can be further shown that for any choices of $a(x)$ and $b(t)$, the equation \eqref{b01} is linearizable by a point transformation, i.e., it admits a maximal Lie algebra of point symmetries isomorphic to $\mathfrak{sl}(3,\mathbb{R})$ \cite{Mah}.\footnote{In Appendix A, the symmetry generators are explicitly given.} However, as the point transformation is not canonical, it may change the physical meaning of the dependent and independent variables, and thus this does not constitute the most appropriate strategy for applications. On the other hand, the first-order equation \eqref{b03} may not always be explicitly integrable, in the sense that $x$ can be expressed in terms of elementary functions of the independent variable $t$.

As already mentioned, in various physical applications the use of \eqref{b02} may  not be  computationally satisfactory, and an alternative approach to derive exact solutions (whenever possible) is required. In this context, it was observed in \cite{LH2015,BHLS} that the equation \eqref{a1}, when interpreted as a first-order system of ODEs, possesses the supplementary structure of a Lie--Hamilton (LH in short) system associated to the so-called `book algebra' $\mathfrak{b}_2$ (see \cite{
LH2015, BHLS, LuSa, Bernoulli} and references therein). It was further shown in \cite{Bernoulli} that any LH system based on this Lie algebra can be integrated by quadratures, a property that also holds, under certain constraints, for its quantum deformation~\cite{ECH}.

 The structure of the paper is the following. In Section~\ref{s2} we reconsider the generalized Buchdahl equation \eqref{b01} from the point of view of LH systems, determining the general solution by means of the associated book Lie algebra $\mathfrak{b}_2$, following the general ansatz proposed in \cite{Bernoulli}. The procedure is illustrated through some special types of the generalized Buchdahl equations in Section~\ref{s3}. In particular, it is shown that no further extension of the generalized Buchdahl equation as a LH system based on $\mathfrak{b}_2$ is possible.     In Section~\ref{s4} we study the Poisson--Hopf deformation of the generalized Buchdahl equations associated with the quantum book algebra~\cite{ECH}, showing that the corresponding equations can also be solved exactly, and  then apply such results in Section \ref{s5} to the special cases considered before in Section \ref{s3}, with special emphasis on the first-order approximation in the quantum deformation parameter $z$. It is shown that these equations can also be solved exactly.   In Section \ref{s6}, we extrapolate the results so obtained to the oscillator algebra $\mathfrak{h}_4$ extending $\mathfrak{b}_2$, finding that the corresponding systems can still be solved exactly. The procedure is illustrated considering the extension of the special cases studied in Section \ref{s3}. In Section \ref{s7}, their Poisson--Hopf deformation, based on the `non-standard' quantum deformation of  $\mathfrak{h}_4$, is analyzed, also obtaining its general solution that is a completely novel result. In addition, it is 
   applied to the systems in Section~\ref{s5} arriving at their extended  version. In Section \ref{s8}, the physical and mathematical significance of the quantum deformation parameter $z$ is considered in more detail. Specifically, it is shown that for higher-dimensional deformations, the corresponding Hamiltonian equations exhibit interaction terms that are intrinsic to the deformation, meaning that the equations of the motion are no longer copies of the original one-dimensional system. Finally, in a concluding section we summarize the results obtained concerning (deformed) generalized Buchdahl equations based on $\mathfrak{b}_2$ and $\mathfrak{h}_4$,  also highlighting how these systems are related by a limiting process. Moreover, the existence of a Lagrangian formalism for any (quantum deformed) equation obtained from \eqref{a1} is indicated. Some conclusions are drawn, and possible future developments of the method in connection with the embedding of the book algebra into either the simple Lie algebra $\mathfrak{sl}(2,\mathbb{R})\supset \mathfrak{b}_2$ or the two-photon algebra $\mathfrak{h}_6\supset\mathfrak{h}_4\supset \mathfrak{b}_2$  are briefly discussed.

 
 \section{The generalized Buchdahl equation as a Lie--Hamilton system}
 \label{s2}
   
 We convene to call \eqref{b01} the generalized Buchdahl equation, as considered in~\cite{Nikiciuk}. For the special choice $a(x)=3/x$ and $b(t)=1/t$, the classical Buchdahl equation is recovered. As shown in \cite{Nikiciuk}, \eqref{b01} admits both a standard and reciprocal Lagrangian description for arbitrary choices of the parameter functions. The generalized equation, besides its applications in the context of General Relativity, can also be interpreted as the equation of the motion for     nonlinear  systems with variable coefficients.

  We first reinterpret the generalized Buchdahl equation in the context of the LH systems framework, deriving its general solution in appropriate coordinates after a canonical transformation, and analyze some relevant particular cases in Section~\ref{s3}. Although this approach does not supply essential new information (the equation being solved by \eqref{b02}), it is relevant for the study of perturbations or quantum deformations of the equation, which are, in general, no longer exact and linearizable by a point transformation, and where the application of the traditional direct integration methods may be too cumbersome.

Considering the variable $y\equiv \dd x/\dd t $, the equation \eqref{b01} is equivalent to the first-order nonlinear system of ODEs
  \begin{equation}
  \frac{\dd x}{\dd t}=y,\qquad
 \frac{\dd y}{\dd t}=a(x)y^2+b(t)y .
 \label{b2}
 \end{equation}
 This system can further be expressed in terms of the $t$-dependent vector field
 \be
 \>X_t=\>X_1+b(t)\>X_2,
  \label{b3}
 \ee 
 where the vector fields $\>X_1$ and $\>X_2$ are explicitly given by
 \begin{equation}
 \>X_1=y\frac{\partial }{\partial x}+a(x)y^2\frac{\partial}{\partial y}\, ,\qquad  \>X_{2}=y\dfrac{\partial}{\partial y}\, .
 \label{b4}
 \end{equation}
These vector fields satisfy the Lie bracket 
\be
 [\>X_2,\>X_1]=\>X_1,
  \label{b5}
\ee
hence they span a Lie algebra isomorphic to the book algebra $\mathfrak{b}_2$.\footnote{This is the same as the affine algebra in two dimensions.}
The generator $\>X_2$ can be seen as a dilation, while $\>X_1$ corresponds to a translation. Formally, (\ref{b2}) and (\ref{b3}) define a Lie system~\cite{LSc,PW,CGM00} with a non-invariance Lie algebra (called Vessiot--Guldberg Lie algebra) isomorphic to the `book'  Lie algebra $\mathfrak{b}_2$.   It can actually be shown (see e.g. \cite{C132}) that the ODE \eqref{b01} is the most general second-order scalar equation that admits $\mathfrak{b}_2$ as a Vessiot--Guldberg algebra.  However, the generalized Buchdahl equation is more than a mere Lie system, as it admits a symplectic form $\omega$ such that the vector fields $\>X_i$  are Hamiltonian vector fields with respect to appropriate Hamiltonian functions $h_i$ $(i=1,2)$. The compatibility conditions 
\be
\mathcal{L}_{\>X_i}\omega=0,\qquad \iota_{{\bf X}_i}\omega={\rm d}h_i  ,
\label{b6}
\ee
with $\iota$ denoting the contraction or inner product of $\omega$, determine the structure of a LH system (see \cite{LuSa} and references therein).\footnote{According to the classification of LH systems on $\mathbb R^2$~\cite{LH2015,BHLS}, the  book Lie algebra $\mathfrak{b}_2$   is locally diffeomorphic to the imprimitive class ${\rm I}_{14A}$ with $r=1$ and $\eta_1(x)={\rm e}^x$.} 

In this case, the symplectic form is given by  
\begin{equation}
\omega=\dfrac{\exp\left(-\int^x a(\xi)\dd \xi\right)}{y}\,\dd x\wedge \dd y, 
  \label{b7}
\end{equation}
and hereafter it will be assumed  that $y\ne 0$.  The Hamiltonian functions associated to the Hamiltonian vector fields $\>X_i$ are  
\be
h_1=y\exp\left(-\int^x a(\xi)\dd \xi\right),\qquad h_2=-\int^x\exp\left(-\int^\xi a(  \xi')\dd  \xi'\right)\dd \xi ,
  \label{b9}
\ee
which satisfy the following Poisson bracket with respect to $\omega$ (see (\ref{b5})):
\be \{h_{2},h_{1}\}_\omega=-h_{1}.
\nonumber
\ee
Hence, the LH system (\ref{b3}) has a $t$-dependent Hamiltonian given by
\be
h_t=h_{1}+b(t) h_{2},
  \label{b11}
\ee
whose Hamilton equations  with respect to the non-canonical symplectic form $\omega$ (\ref{b7}) give rise to the equations (\ref{b2}):
 \be
\begin{split}
\frac {\dd x}{\dd t}&=\{x, h_t\}_\omega = \frac{y}{\exp\left(-\int^x a(\xi)\dd \xi\right)}\, \frac{\partial h_t}{\partial y} , \\[4pt]
\frac {\dd y}{\dd t}&=\{y, h_t\}_\omega = -\frac{y}{\exp\left(-\int^x a(\xi)\dd \xi\right)}\, \frac{\partial h_t}{\partial x}  .
 \end{split}
\nonumber
\ee
We observe that all the above structures are properly defined on $\mathbb R^2_{y\neq 0}$.


\subsection{The general solution as a LH system}
\label{s21}
 
LH systems related to the book algebra were generically studied in~\cite{LH2015}, with nonlinear superposition rules being given in \cite{BHLS}, using an appropriate embedding of $\mathfrak{b}_2$ into a higher-dimensional LH algebra. The generalized Buchdahl equation is one particular case of an ample class of 
$\mathfrak{b}_2$-LH systems, that comprise, among others, complex Bernoulli differential equations (with $t$-dependent real coefficients) \cite{Bernoulli}, some Lotka--Volterra systems as well as various types of time-dependent epidemic models with stochastic fluctuations \cite{CFH}, for which an exact solution can be found using 
an explicit diffeomorphism (corresponding to an appropriate change of variables) that leads to a canonical realization of the LH algebra (see \cite{Bernoulli,CFH} for further details).
 
As  a shorthand notation,  we define the function
 \be
 \xii(x):=\exp\left(-\int^x a(\xi)\dd \xi\right) ,
\nonumber
\ee
so that the expressions (\ref{b7}) and (\ref{b9}) read
\be
\omega=\dfrac{\xii(x)}{y}\,\dd x\wedge \dd y ,\qquad h_1=y\, \xii(x),\qquad h_2=-\int^x \xii(\xi)\,\dd \xi \, .
 \label{c2}
\ee
 Now, under the change of variables
 \be
 q=-y\,  \xii(x),\qquad p=\frac 1 y \,{  \xii(x)^{-1} {\int^x  \xii(\xi) \dd \xi}}\, ,
 \label{c3}
 \ee
the Buchdahl equation, as presented above, can be reformulated in terms of the $\mathfrak{b}_2$-LH algebra in the canonical form introduced in~\cite{Ballesteros6} (see also~\cite{CFH,Bernoulli})), which differs considerably from that given in~\cite{LH2015,BHLS}, although both are (locally) diffeomorphic.
 In this context, the coordinates $(q,p)$ are canonical, and the symplectic form (\ref{b7})  and Hamiltonian functions (\ref{b9})  become
\be
\omega\equiv \wcan = \dd q\wedge  \dd p,
\label{c4}
\ee
 \be
 h_1=  - q , \qquad h_2=  qp,\qquad \{ h_2,h_1\}_\wcan= - h_1 ,
\label{c5}
\ee
while the $t$-dependent Hamiltonian (\ref{b11}) reduces to 
\begin{equation} 
 h_t= h_1 + b(t)  h_2     = - q+  b(t)  qp.  
 \label{c6}
\end{equation}
The corresponding equations of the motion are  given by
  \begin{equation} 
\frac{\dd q}{\dd t}=\{q, h_t\}_\wcan=b(t) q ,
\qquad
\frac{\dd p}{\dd t}=\{p, h_t\}_\wcan=1-b(t) p  ,
 \label{c7}
\end{equation}
which is a linear, uncoupled system. We observe that $a(x)$ does not appear explicitly in (\ref{c7}) (compare with (\ref{b2})), as it is `hidden' within the transformation (\ref{c3}) through $\xii(x)$. The Hamiltonian vector fields (\ref{b4}), when expressed in the canonical variables, are 
 \be
  \>X_1=\frac{\partial}{\partial p}\, , 
 \qquad     \>X_2=q\, \frac{\partial}{\partial q}- p\, \frac{\partial}{\partial p}\, ,   \label{c8}    
\ee
 which satisfy the Lie commutator (\ref{b5}) and provide   the  same LH system (\ref{c7}). As follows from (\ref{c7}), the generalized Buchdahl LH system (\ref{b2})     is separable in the coordinates $(q,p)$, and can be easily solved by quadratures: 
  \be
\begin{split}
q(t)&=c_1 \,\eee^{\gam(t)},\qquad \gam(t):= \int^t b(\tau)\dd \tau ,\\[4pt]
p(t)&=\left( c_2 +  \int^t \eee^{\gam(\tau )}  \dd \tau\right)  \eee^{-\gam(t)},
\end{split}
\label{c9}
\ee
where $c_1$ and $c_2$ are the two  constants of integration determined by the  initial conditions.
 These results are summarized in the following statement.

 \begin{proposition}
 \label{prop1}
The general solution of the generalized second-order  Buchdahl equation (\ref{b01}) and its associated first-order system of ODEs (\ref{b2}) is given by
  \be
\begin{split}
&\int^x \xii(\xi)\dd \xi = - c_1 \left( c_2 +  \int^t \eee^{\gam(\tau)}  \dd \tau\right)  , \qquad  \xii(x):=\exp\left(-\int^x a(\xi)\dd \xi\right), \\[4pt]
&y(t)=\frac{\dd x}{\dd t}= -c_1 \, \frac{\eee^{\gam(t)}  }{ \xii(x)} \,   ,\qquad \gam(t):= \int^t b(\tau)\dd \tau ,
 \end{split}
\label{c10}
\ee
where $c_1$ and $c_2$ are the two  constants of integration determined by the  initial conditions.
 \end{proposition}

Observe in particular that the first equation of (\ref{c10}) can also be expressed as
\be
\frac{\dd }{\dd t}\left(  \eee^{-\gam(t)}\frac{\dd }{\dd t}    \int^x \xii(\xi)\dd \xi         \right) =0 \quad \equiv \quad 
\frac{\dd }{\dd t}\left(  \eee^{-\gam(t)} \, \xii(x)  \,\frac{\dd x}{\dd t}     \right) =0,
\label{c10a}
\nonumber
\ee
which, taking into account that
\be
\frac{\dd }{\dd t} \, \eee^{-\gam(t)} = -\eee^{-\gam(t)} b(t)  ,\qquad \frac{\dd }{\dd x}\,\xii(x) = -  \xii(x)a(x)
 ,\qquad \frac{\dd }{\dd t}\,\xii(x) = -  \xii(x)a(x)\,\frac{\dd x}{\dd t} \, ,
\label{c10b}
\nonumber
\ee
  reproduces the initial generalized second-order  Buchdahl equation (\ref{b01}).

In the context of   nonlinear  dynamical systems with variable coefficients, the previous result can be applied directly by means of appropriate choices for the parameter functions $a(x)$ and $b(t)$. In this situation, the functions $\xii$ and $\gam$ are obtained from (\ref{c10}), reducing the problem to the integration of $x(t)$; while $y(t)$ follows directly  by insertion of $x(t)$ into the second equation, providing the solution of system  (\ref{b2}).  

 
  \section{Applications to   particular generalized Buchdahl equations}
  \label{s3}
 
We illustrate the procedure in terms of LH systems, by studying some particular cases corresponding to explicit choices for the  parameter  functions $a(x)$  and $b(t)$.
 
 
  \subsection{The classical Buchdahl equation}
  \label{s31}
  
The equation arises from the specific choices 
\be
a(x)=\frac 3 x\, ,\qquad b(t)=\frac 1 t \, ,\qquad x\in  \mathbb R^\ast, \qquad t\in  \mathbb R^\ast
\label{c11}
\ee
in (\ref{b2}), 
so that the functions $\xii(x)$ and $\gam(t)$ in (\ref{c10}) are given by 
\be
    \xii(x)=x^{-3} ,\qquad \gam(t )=\ln t .
\nonumber
\ee
Note that $\eee^{\gam(t)} = t$  appearing in (\ref{c10}) is therefore always well-defined.
It follows that the symplectic form and Hamiltonian functions (\ref{c2}) are given by
\be
\omega=\dfrac{ 1}{x^3y}\,\dd x\wedge \dd y ,\qquad h_1=\frac{y}{x^3}\, ,\qquad h_2=\frac{1}{2x^2}\, .
\nonumber
\ee
Substituting these expressions into the first equation of (\ref{c10}), we find that
\be
\frac{1}{x^2(t)} = 2 c_1c_2+ c_1  {t^2},
\nonumber
\ee
from which the general solution 
\be
x(t)=\frac{\pm 1}{ \sqrt{2 c_1c_2 +c_1 t^2} }\,  ,\qquad y(t)= \frac{\mp c_1 t}{  \bigl(2 c_1c_2 +c_1 t^2\bigr)^{3/2} }\, 
 \label{c15}
\ee
is obtained, recovering the  expression (\ref{a2}) of   the Buchdahl equation (\ref{a1}) with $c_1= k_1^2k_2$, $c_2=1/(2 k_2)$ and $f(r)\equiv x(t)$.

\subsection{Case with $a(x)=1/x$ and arbitrary $ {b(t)} $}
\label{s32}
  
As a first natural generalization, we consider in (\ref{b2}) the function $a(x)=x^{-1}$ and an arbitrary $b(t)$. It follows that  
\be
    \xii(x)=x^{-1} , \qquad x> 0,
\nonumber
\ee
with $\gam(t)$ given in (\ref{c10}). The symplectic form and Hamiltonian functions (\ref{c2}) read
 \be
\omega=\dfrac{ 1}{xy}\,\dd x\wedge \dd y ,\qquad h_1=\frac{y}{x}\, ,\qquad h_2=-\ln x .
\nonumber
\ee
The first equation in (\ref{c10}) provides 
\be
\ln x(t) = - c_1 \left( c_2 +  \int^t \eee^{\gam(\tau)}  \dd \tau\right),
\nonumber
\ee
which can be easily solved, leading to the exact solution  
 \be
\begin{split}
x(t)=\exp\left(- c_1c_2 -c_1  \int^t \eee^{\gam(\tau)}  \dd \tau \right)  ,\quad
 y(t)=-c_1\, \eee^{\gam(t) }\exp\left(- c_1c_2 -c_1  \int^t \eee^{\gam(\tau)}  \dd \tau  \right).
 \end{split}
\nonumber
\ee
For instance, choosing $b(t)=1/t$  with $t\in  \mathbb R^\ast$    (i.e.~$\eee^{\gam(t)} = t$), we find that
\be
x(t)=\exp\left(- c_1c_2 -c_1  \frac{t^2}2\right)  ,\qquad y(t)=-c_1 t\exp\left(- c_1c_2 -c_1  \frac{t^2}2\right),
 \label{c20}
\ee
to be compared with (\ref{c15}).

 
\subsection{Case with $  a(x)=\aaa/x$  
 $(\aaa\ne 1)$ and arbitrary $b(t)$ }
\label{s33}

We now choose $a(x)=\aaa x^{-1}$ with $\aaa\in  \mathbb R^\ast$ and $\aaa\ne 1$, keeping $b(t)$ arbitrary, so that 
\be
    \xii(x)=x^{-\aaa}, \qquad x\in  \mathbb R^\ast .
\nonumber
\ee
Hence the expressions (\ref{c2}) give rise to
 \be
\omega=\dfrac{ 1}{x^\aaa y}\,\dd x\wedge \dd y ,\qquad h_1=\frac{y}{x^\aaa}\, ,\qquad h_2=-\frac{x^{1-\aaa}}{1-\aaa}\,  , \qquad \aaa\ne 1.
\nonumber
\ee
The first equation in (\ref{c10}) reads
\be
\frac{x^{1-\aaa}(t)}{1-\aaa}= - c_1 \left( c_2 +  \int^t \eee^{\gam(\tau)}  \dd \tau\right)   ,
\nonumber
\ee
providing the following exact solution 
  \be
\begin{split}
x(t)&= \left( (\aaa-1) c_1\left( c_2 +  \int^t \eee^{\gam(\tau)}  \dd \tau \right) \right)^{\frac{1}{1-\aaa} } ,\\[4pt]
 y(t)&=-c_1\, \eee^{\gam(t) }   \left( (\aaa-1) c_1\left( c_2 +  \int^t \eee^{\gam(\tau)}  \dd \tau \right) \right)^{\frac{\aaa}{1-\aaa} }.
 \end{split}
\nonumber
\ee
As a particular case, taking $b(t)=1/t$ with $t\in  \mathbb R^\ast$,  we find that
\be
x(t)= \left( (\aaa-1)c_1 \biggl( c_2 +\frac{t^2}2\biggr) \right)^{\frac{1}{1-\aaa} } ,\qquad y(t)=- c_1 t\left( (\aaa-1)c_1 \biggl( c_2 +\frac{t^2}2\biggr) \right)^{\frac{\aaa}{1-\aaa} } ,
\nonumber
\ee
which, as expected, reduces for $\aaa=3$ to the solution (\ref{c15}) of the classical Buchdahl equation.

 
  \subsection{Case with $a(x)=\aaa x^\rr$ $(\rr\ne -1)$ and arbitrary $b(t)$}
  \label{s34}

Despite the previous particular cases determined by Proposition~\ref{prop1}, for which exact solutions have been presented in an explicit form, it is worth observing that, in general, the first equation in (\ref{c10}) does not provide an explicit expression for $x(t)$, as it may not be expressible in terms of the usual elementary functions.

The simplest choice for which this pattern is observed corresponds to polynomial choices of $a(x)$, more precisely 
\be
a(x)=\aaa x^\rr ,\qquad x\in  \mathbb R^\ast,\qquad  \aaa\in  \mathbb R^\ast,\qquad \rr\in\mathbb R^\ast,\qquad \rr\ne -1 .
\nonumber
\ee
Then we find that
\be
    \xii(x)=\exp\biggl(-\aaa\, \frac{x^{\rr+1}}{\rr+1} \biggr) .
\nonumber
\ee
along with
 \be
\begin{split}
\omega&=\dfrac{ 1}{ y}\, \exp\biggl(-\aaa\, \frac{x^{\rr+1}}{\rr+1} \biggr) \dd x\wedge \dd y ,\qquad h_1= {y} \exp\biggl(-\aaa\, \frac{x^{\rr+1}}{\rr+1} \biggr)\, ,\\[4pt]
 h_2&=\frac{1}{\rr+1}\, \biggl( \frac{\aaa}{\rr+1} \biggr)^{-\frac{1}{\rr+1} } \Gamma\biggl(\frac{1}{\rr+1} ,  \aaa\, \frac{x^{\rr+1}}{\rr+1}  \biggr), \qquad \rr\ne -1,
  \end{split}
\nonumber
\ee
where $\Gamma(u, v)$ denotes the incomplete Gamma function \cite{AS}. The first equation in (\ref{c10}), that provides the general solution $x(t)$, adopts the cumbersome implicit form
\be\frac{1}{\rr+1}\, \biggl( \frac{\aaa}{\rr+1} \biggr)^{-\frac{1}{\rr+1} } \Gamma\biggl(\frac{1}{\rr+1} ,  \aaa\, \frac{x^{\rr+1}}{\rr+1}  \biggr) =  c_1 \left( c_2 +  \int^t \eee^{\gam(\tau)}  \dd \tau\right)   .
\nonumber
\ee
For general values of $\alpha$, this expression cannot be solved explicitly with respect to $x(t)$, and alternative methods, like the Lie series  \cite{Groe}, have to be applied to determine the solution of the system.

 
  \subsection{Non-existence of $\mathfrak{b}_2$-based extensions of the generalized Buchdahl equation}
  \label{s35}

Taking into account the  $\mathfrak{b}_2$-LH algebra symmetry (see~\eqref{b2}--\eqref{b11}) of the generalized Buchdahl system, it is rather natural to analyze whether it is possible to extend the system adding a second $t$-dependent arbitrary function, i.e., to consider two coefficient functions $b_1(t)$ and $b_2(t)$. We start with the first-order ODE system  
  \begin{equation}
  \frac{\dd x}{\dd t}=b_1(t) y,\qquad
 \frac{\dd y}{\dd t}=b_1(t)a(x)y^2+b_2(t)y ,
 \label{c30}
 \end{equation}
such that $b_1(t)$ and $b_2(t)$ are arbitrary. These equations determine a Lie system with the $t$-dependent vector field $\>X_t=b_1(t)\>X_1+b_2(t)\>X_2$, where the $\>X_i$ are defined as in (\ref{b4}). It is straightforward to verify that (\ref{c30}) also determines a $\mathfrak{b}_2$-LH system with Hamiltonian $h_t=b_1(t) h_{1}+b_2(t) h_{2}$ and the same Hamiltonian functions (\ref{b9}) and symplectic form (\ref{b7}).

Assuming that $b_1(t)\neq 0$, we consider the following change of coordinates in $y$:
\be
\tilde y(t)= b_1(t) y(t). 
\nonumber
\ee
It follows that 
\be
\frac{\dd \tilde y}{\dd t}=b_1\, \frac{\dd  y}{\dd t}+ \frac{\dd   b_1}{\dd t}\, y=
b_1^2 a(x) y^2 + b_1b_2 y +  \frac{\dd   b_1}{\dd t}\, y= a(x) \tilde y^2+ b(t)\tilde y,
\nonumber
\ee
where 
\be
b(t)=b_2(t)+\frac 1{b_1(t)}  \frac{\dd   b_1}{\dd t}. 
\nonumber
 \ee
Then, the system \eqref{c30} is equivalent to \eqref{b2}, showing that the latter   cannot be generalized as a LH-system based on the LH algebra $\mathfrak{b}_2$,   in consistence with the results derived in \cite{C132}.


\section{Deformed generalized Buchdahl    equation from the quantum\\  book algebra}
  \label{s4}
  
Starting from the quantum deformation of the book algebra, which is briefly recalled, in this section we obtain the deformed counterpart of the generalized Buchdahl system (\ref{b2}), together with its corresponding general solution, thus extending Proposition~\ref{prop1} to an arbitrary quantum deformation parameter $z$, while keeping the parameter functions $b(t)$ and $a(x)$ arbitrary.


\subsection{Quantum  book algebra and deformed book LH systems}
  \label{s41}

The quantum algebra deformation (see~\cite{CP,Abe} for details) of the  book Lie algebra $\mathfrak{b}_2$ is denoted by $U_{   z}(\mathfrak{b}_2)\equiv \mathfrak{b}_{z,2 }$, where $z$ is the quantum deformation   real parameter (the usual  $q=\eee^z$). Its Hopf structure is defined by the following deformed coproduct map $\Delta_z$ and   compatible commutation relation in an `abstract' basis $(v_1,v_2)$:
 \be
\begin{split}
\Delta_z(v_1)&=v_1\otimes 1 + 1 \otimes v_1 , \\[2pt]
\Delta_z(v_2)&=v_2\otimes \eee^{-z v_1}+1\otimes v_2 , \\[1pt]
[v_2,v_1]_z&=- \frac{1-\eee^{-z v_1}}{z} \, ,
\end{split}
\label{d1}
\ee
such that $\Delta_z$ is an algebra homomorphism and   satisfies the coassociativity condition 
\be
({\rm Id}\otimes\Delta_z)\Delta_z =(\Delta_z\otimes {\rm Id})\Delta_z .
\nonumber
\ee
The pair $(\mathfrak{b}_{z,2 },\Delta_z)$ hence defines a coalgebra structure (see \cite{Bernoulli,ECH} and references therein for more details). In the following,  
we apply to $\mathfrak{b}_2$ the formalism of Poisson--Hopf deformations of LH systems introduced in~\cite{Ballesteros6,BCFHL}, from which a   deformation of the generalized Buchdahl system (\ref{b2}) will be derived.

A deformed  symplectic representation $D_z$ of $\mathfrak b_{z,2}$  (\ref{d1}) in terms of  the canonical variables $(q,p)$ of Section~\ref{s21} and the canonical symplectic form (\ref{c4}) is given by
 \be
 h_{z,1}:= D_z(v_1)=  -q , \qquad  h_{z,2}:= D_z(v_2)=  \left( \frac{ \eee^{z q} -1}{z}\right) p,
\label{d3}
\ee
where the corresponding deformed Poisson bracket with respect to $\wcan$ is
\be
\{h_{z,2},h_{z,1}\}_\wcan=\frac{\eee^{-z h_{z,1}} -1}{z}\, .
\nonumber
\ee
From the relation $\iota_{{\bf X}_{z,i}}\wcan={\rm d}h_{z,i}$, we compute   the corresponding deformation of the vector fields (\ref{c8})
\be
 \>X_{z,1}= \frac{\partial}{\partial p} \, ,   \qquad  \>X_{z,2}=\left( \frac{ \eee^{z q}-1}{z} \right)\frac{\partial}{\partial q}- \eee^{z q} p \, \frac{\partial}{\partial p} \, ,
\nonumber
\ee
 which span a smooth distribution in the sense of  Stefan--Sussmann~\cite{Vaisman,Pa57,WA} through the commutator
 \be
 [{\bf X}_{z,2},{\bf X}_{z,1}]=  \eee^{z q} \, {\bf X}_{z,1}.
\nonumber
 \ee
The invariance condition of $\wcan$ in (\ref{c4}) under the Lie derivative (\ref{b6}) is trivially satisfied. 

This leads to the deformed $t$-dependent Hamiltonian and $t$-dependent vector field (compare with (\ref{c6}))
\begin{equation} 
\begin{split}
h_{z,t}&=h_{z,1} + b(t)  h_{z,2}    =  -  q  + b(t)   \biggl( \frac{ \eee^{z q} -1}{z}\biggr) p , \\[2pt]
  \>X_{z,t}&=\>X_{z,1}  + b(t)   \>X_{z,2}   =  \frac{\partial}{\partial p}+ b(t) \left( \left( \frac{ \eee^{z q}-1}{z} \right)\frac{\partial}{\partial q}- \eee^{z q}p \, \frac{\partial}{\partial p} \right)      ,
\end{split}
\nonumber
\end{equation}
where $b(t)$ is an    arbitrary real parameter function. The associated first-order system of non-autonomous nonlinear and coupled  ODEs on $\mathbb R^2$  is given by
\begin{equation} 
\begin{split}
\frac{\dd q}{\dd t}&=  b(t) \biggl( \frac{ \eee^{z q}-1}{z} \biggr)   , \\[2pt]
\frac{\dd p}{\dd t}&=1 - b(t) \eee^{z q} p  . 
\end{split}
 \label{d8}
\end{equation}
For the limit $z\rightarrow 0$, the system (\ref{c7}) is recovered.\footnote{All the above deformed expressions reduce to (\ref{c5})--(\ref{c9}) concerning the classical   $\mathfrak{b}_2$-LH algebra.} As the first of the equations above is separable, it can be easily solved, and substitution into the second equation yields, after a quadrature, the general solution (see \cite{Bernoulli,ECH})
\be
\begin{split}
q(t)&=-\frac{\ln \left(1- z c_1 \eee^{\gam(t)} \right)}z   ,\qquad \gam(t):= \int^t b(\tau)\dd \tau ,\\[4pt]
p(t)&=   \left( \eee^{-\gam(t)} -z c_1   \right)  \left( c_2 +  \int^t    \frac{1}{\eee^{-\gam(\tau)}-z c_1 } \,\dd \tau \right)  ,
\end{split}
\label{d9}
\ee
where  $c_1$ and $c_2$ are the two  constants of integration determined by the  initial conditions. Observe that the first equation in (\ref{d9}) can be expressed alternatively as
\be
\frac{1- \eee^{-zq(t)} }{z}= c_1 \,\eee^{\gam(t)}.
\nonumber
\ee

It is worthy to be observed  that the presence of the quantum deformation parameter $z$  can be regarded as the introduction of a perturbation in  the  classical  $\mathfrak{b}_2$-LH system (\ref{c7}), in such a manner that  a nonlinear interaction or coupling between the  variables $(q,p)$ in the deformed $\mathfrak{b}_2$-LH system (\ref{d8})   arises  through the term $   \eee^{z q} p$. This fact can be clearly appreciated 
by taking a power series expansion in $z$ of  (\ref{d8}) and truncating at the first-order, leading to the system 
\begin{equation} 
\begin{split}
\frac{\dd q}{\dd t}&= b(t)  \bigl(  q  +\tfrac 12 z    q^2 \bigr) +o[z^2]  , \\[2pt]
\frac{\dd p}{\dd t}&= 1 - b(t) \bigl(   p +z     qp  \bigr)+o[z^2]   ,
\end{split}
 \label{d11}
\end{equation}
which hold for a small value of   $z$. In this approximation, we find that   $z$ introduces a quadratic term $q^2 $ in the first equation of  (\ref{d11}), leading to a real Bernoulli equation, while the second equation is linear in $p$, once the value of $q$ has been obtained. This enables to integrate the system by quadratures, allowing us to obtain the general solution.


\subsection{Deformed generalized Buchdahl  equation and its general solution}
  \label{s42}
  
By introducing the change of variables (\ref{c3}) into the expressions of (\ref{d3}) to (\ref{d8}) with the initial canonical variables $(q,p)$, we  obtain directly the 
deformation of the  generalized Buchdahl equation presented in Section~\ref{s2} in the appropriate coordinates $(x,y)$. This result is summarized as follows.

 \begin{proposition}
 \label{prop2}
(i)  The deformation of the $t$-dependent generalized Buchdahl Hamiltonian (\ref{b11}) in terms of the variables $(x,y)$ is defined by
\begin{equation} 
\begin{split}
h_{z,t}&=h_{z,1} + b(t)  h_{z,2}   , \\[2pt]
h_{z,1} & =y\, \xii(x)   ,\qquad \xii(x):=\exp\left(-\int^x a(\xi)\dd \xi\right) ,\\[2pt]
h_{z,2} & =\frac{\exp\bigl(-z y\, \xii(x)\bigr) -1}{z y\, \xii(x)}\int^x \xii(\xi)\,\dd \xi  \, , 
\end{split}
 \label{d12}
\end{equation}
such that
\be
\{h_{z,2},h_{z,1}\}_\omega=\frac{ \eee^{-z h_{z,1}} -1} {z}\, 
\nonumber
\ee
with respect to the non-canonical symplectic form $\omega$ (\ref{c2}).

\noindent
(ii) The deformation of the  generalized Buchdahl system (\ref{b2})  is given by
\begin{equation} 
\begin{split}
\frac {\dd x}{\dd t}&=\{x, h_{z,t}\}_\omega = y + b(t) \left( \frac{\exp\bigl(z y\, \xii(x)\bigr) -1 -z y\,\xii(x) }{z y\,\xii^2(x)}\right)  \exp\bigl(-z y\, \xii(x)\bigr) \int^x \xii(\xi)\,\dd \xi  \, ,\\[2pt]
\frac {\dd y}{\dd t}&=\{y, h_{z,t}\}_\omega = a(x) y^2 + b(t)  \exp\bigl(-z y\, \xii(x)\bigr) \\[2pt]
&\qquad\qquad  \times \left(     \frac{\exp\bigl(z y\, \xii(x)\bigr) -1}{z \, \xii(x)} + \frac{ \exp\bigl(z y\, \xii(x)\bigr) -1 -z y\,\xii(x) }{z \,\xii^2(x)}\, a(x) \int^x \xii(\xi)\,\dd \xi
  \right)  ,\end{split}
 \label{d14}
\end{equation}
for arbitrary $a(x)$, $b(t)$ and $z$.
 \end{proposition}

It follows that the introduction of the   Poisson--Hopf  deformation of the book algebra  leads to  the appearance of strong   nonlinear interaction terms in the initial generalized Buchdahl system (\ref{b2}) determined by the quantum deformation parameter $z$. In this sense, note that  the first equation in (\ref{d14}) is no longer equal to  $ {\dd x}/{\dd t}=y$ and additional functions depending on $(x,y)$ and $z$, as well as the coefficient  $b(t)$ itself, enter into the deformation. At the first-order approximation in $z$ the system  (\ref{d14})  reduces to
\begin{equation} 
\begin{split}
\frac {\dd x}{\dd t}&=  y + \frac 12 z y \, b(t)  \int^x \xii(\xi)\,\dd \xi +o[z^2]  ,\\[2pt]
\frac {\dd y}{\dd t}&=  a(x) y^2 + b(t) y \left( 1- \frac 12 z y \biggl( \xii(x)-  a(x) \int^x \xii(\xi)\,\dd \xi \biggr)  \right)  +o[z^2]  , \end{split}
 \label{d15}
\end{equation}
which couples nontrivially the coordinate functions.

 The deformed vector fields corresponding to the Hamiltonian (\ref{d12}) are given by 
 \begin{equation} {\small
\begin{split}
 \>X_{z,t}&=\>X_{z,1}+b(t)\>X_{z,2}  ,\qquad 
 \>X_{z,1}=y\frac{\partial }{\partial x}+a(x)y^2\frac{\partial}{\partial y}\, ,\\[2pt]
 \>X_{z,2}&=\left( \frac{\exp\bigl(z y\, \xii(x)\bigr) -1 -z y\,\xii(x) }{z y\,\xii^2(x)}\right)  \exp\bigl(-z y\, \xii(x)\bigr) \int^x \xii(\xi)\,\dd \xi \, \dfrac{\partial}{\partial x}\\[2pt]
 &\qquad +    \exp\bigl(-z y\, \xii(x)\bigr)  \left(     \frac{\exp\bigl(z y\, \xii(x)\bigr) -1}{z \, \xii(x)} + \frac{ \exp\bigl(z y\, \xii(x)\bigr) -1 -z y\,\xii(x) }{z \,\xii^2(x)}\, a(x) \int^x \xii(\xi)\,\dd \xi
  \right)\dfrac{\partial}{\partial y}\,   ,
  \end{split}
\nonumber\nonumber }
\end{equation}
 which fulfill the relation (\ref{b6}) with respect to $\omega$ in (\ref{b7}) and provide the same  deformed generalized Buchdahl system (\ref{d14}). These vector fields span a distribution with commutator 
\be
 [{\bf X}_{z,2},{\bf X}_{z,1}]=   \exp\bigl(-z y\, \xii(x)\bigr)  {\bf X}_{z,1}.
\nonumber
 \ee
Despite the complicated expressions of the deformed generalized Buchdahl system (\ref{d14}), we emphasize that a general solution can be derived from the exact solution for deformed book LH systems (\ref{d9}) in canonical variables and the change of coordinates (\ref{c3}). This is achieved as follows.
 
\begin{proposition}
  \label{prop3}
The general solution of the  first-order system of deformed generalized Buchdahl equations (\ref{d14})        is given by  
  \be
\begin{split}
&\int^x \xii(\xi)\dd \xi = \frac{\ln \left(1- z c_1 \eee^{\gam(t)} \right)}z   \left( \eee^{-\gam(t)} -z c_1   \right)     \left( c_2 +  \int^t    \frac{1}{\eee^{-\gam(\tau)}-z c_1 } \,\dd \tau \right) , \qquad \\[4pt] 
&y(t)= \frac{\ln \left(1- z c_1 \eee^{\gam(t)} \right)}{z\,  \xii(x)} \,   , \qquad \xii(x):=\exp\left(-\int^x a(\xi)\dd \xi\right) ,\qquad  \gam(t):= \int^t b(\tau)\dd \tau\, ,
 \end{split}
\label{d18}
\ee
where $c_1$ and $c_2$ are the two  integration constants  determined by the  initial conditions.
 \end{proposition}

Notice that, in contrast to Proposition~\ref{prop1}, now $y\ne \dd x/\dd t$ (see (\ref{d14})). As expected, under the limit $z\to 0$ we recover the undeformed/classical solution presented in Proposition~\ref{prop1}. As in the classical case, one has to choose explicit expressions for $a(x)$ and $b(t)$ in the first equation of (\ref{d18})    and try to derive  the solution $x(t)$. From it, the solution $y(t)$ can be deduced from the second equation. As already mentioned, despite the algorithmic procedure, an explicit integration differs from being a trivial task in the general case.
   
 
  \section{Applications to   particular deformed  Buchdahl equations}
  \label{s5}
  
  We illustrate the results established  in Propositions~\ref{prop2} and \ref{prop3} by constructing the 
  deformed counterpart of the particular generalized Buchdahl equations described in Sections~\ref{s31}--\ref{s33}.

 
  \subsection{Deformation of the proper Buchdahl equations}
  \label{s51}

We choose the   functions $a(x)=3x^{-1}$ ($x\in  \mathbb R^\ast$) and $b(t)=t^{-1}$ ($t\in  \mathbb R^\ast$) as in (\ref{c11}), giving rise to $\xii(x)=x^{-3}$ and $\gam(t)=\ln t $. Thus  the symplectic form and the deformed Hamiltonian vector fields (\ref{d12}) read
\be
\omega=\dfrac{ 1}{x^3y}\,\dd x\wedge \dd y ,\qquad h_{z,1}=\frac{y}{x^3}\, ,\qquad h_{z,2}=x\,\frac{1-\exp\bigl(-\frac{zy}{x^3}\bigr)}{2zy} .
\nonumber
\ee
 The deformed system of Buchdahl equations (\ref{d14}) is given by
\begin{equation} 
\begin{split}
\frac {\dd x}{\dd t}&  = y + \frac 1{2t}  \left( x \exp\biggl(-\frac{zy}{x^3}\biggr)- x^4\, \frac{1-\exp\bigl(-\frac{zy}{x^3}\bigr)}{zy} \right)  ,\\[2pt]
\frac {\dd y}{\dd t} & =\frac 3x\, y^2 +\frac 1{2 t}\left( 3 y \exp\biggl(-\frac{zy}{x^3}\biggr)- x^3\,  \frac{1-\exp\bigl(-\frac{zy}{x^3}\bigr)}{z}\right)   
 .\end{split}
 \label{e2}
\end{equation}
The corresponding  exact solution is obtained by application of Proposition~\ref{prop3}. In particular, the first equation of (\ref{d18})  yields directly the exact solution for $x(t)$, namely
   \be
\frac{1}{x^2(t)} = \frac{2(z c_1 t-1)\ln\bigl(1-z c_1 t \bigr)}{zt}  \left( c_2  -  \frac{z c_1 t+\ln\bigl(1-z c_1 t \bigr)}{z^2 c^2_1} \right).
 \label{e3}
\ee
The second equation   gives the solution $y(t)$ in terms of the above result:
\be
y(t)=x^3(t)\, \frac{\ln\bigl(1-z c_1 t \bigr)}{z  }\,  .
\nonumber
\ee
The exact solution (\ref{c15}) of   the proper Buchdahl equation is recovered under the undeformed limit $z\to 0$; the limit of the first factor in (\ref{e3}) leads to $2 c_1$, while the second factor gives $c_2+t^2/2$.

Furthermore,   we  stress that    the quantum deformation parameter  $z$   can be regarded     as a small integrable perturbation parameter similarly to (\ref{d15}). Under such a first-order approximation in $z$, the system (\ref{e2})   reduces to  
\be
\begin{split}
\frac {\dd x}{\dd t}&  \simeq  y -z\,\frac{y}{4 t x^2} \, ,\\[2pt]
\frac {\dd y}{\dd t} & \simeq\frac 3x\, y^2 + \frac y t- z\, \frac{5y^2}{4 t x^3} \,  ,\end{split}
 \label{e5}
\ee
 where, for simplicity,  we omit the term $o[z^2]$ for such approximations. From (\ref{e5}), and taking into account that only terms at most linear in $z$ are considered, it follows that the first-order deformation of the Buchdahl equation (\ref{a1}) adopts the form 
 \be
 \frac{\dd^2 x}{\dd t^2}\simeq\frac 3 x
 \left(\frac{\dd x}{\dd t}\right)^2+\frac 1t \,\frac{\dd x}{\dd t} + z\,\frac{1}{4 t^2 x^2}\,\frac{\dd x}{\dd t}  \, .
  \label{e6}
 \ee
 In contrast to the non-deformed equation, \eqref{e6} only admits one Lie point symmetry
 \begin{equation*}
{\bf Y}= t\frac{\partial}{\partial t}-\frac{x}{2}\frac{\partial}{\partial x},
\end{equation*}
showing that the maximal symmetry has been broken, i.e., the equation is no more linearizable~\cite{Mah}. According to the previous discussion, the  solution of (\ref{e5}) and (\ref{e6}) is given by  
\be
\begin{split}
x(t)&\simeq
\frac{\pm 1}{ \sqrt{2 c_1c_2 +c_1 t^2} } \left( 1+ z\, \frac{c_1 t (6 c_2 -t^2 ) } { 12(2 c_2 + t^2)}   \right)  ,\\[2pt]
y(t) & \simeq \frac{\mp c_1 t}{  \bigl(2 c_1c_2 +c_1 t^2\bigr)^{3/2} }  \left( 1+ z\, \frac{c_1 t (10 c_2 +t^2) } { 4(2 c_2 + t^2)}   \right)  ,\end{split}
\label{ee66}
\ee
to be compared with equation (\ref{c15}).

\begin{figure}[t]
\centering
\includegraphics[scale=0.60]{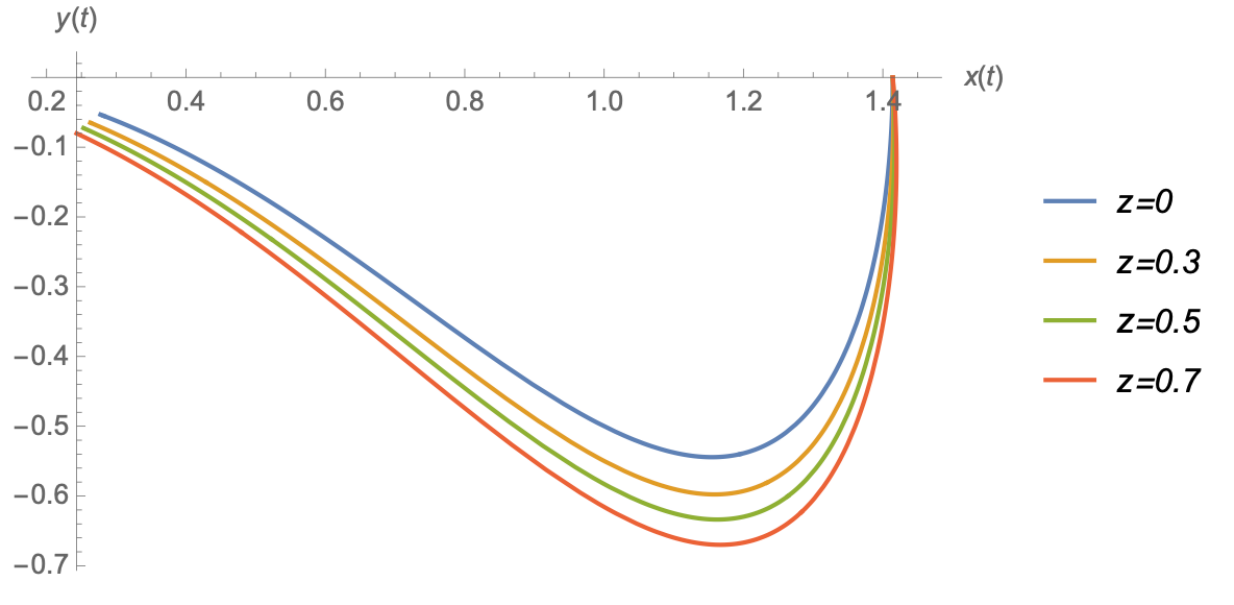}
\caption{\small First-order solutions (\ref{ee66}) of equation (\ref{e6}), for positive $x(t)$ and negative $y(t)$, with the choices of the integration constants $c_1=c_2=0.5$ and positive values of the deformation parameter $z$. The case $z=0$ is  the solution (\ref{c15}) of the proper  Buchdahl equation (\ref{a1}).}
\centering
     \label{fig1}
\end{figure}

\begin{figure}[t]
\centering
\includegraphics[scale=0.6]{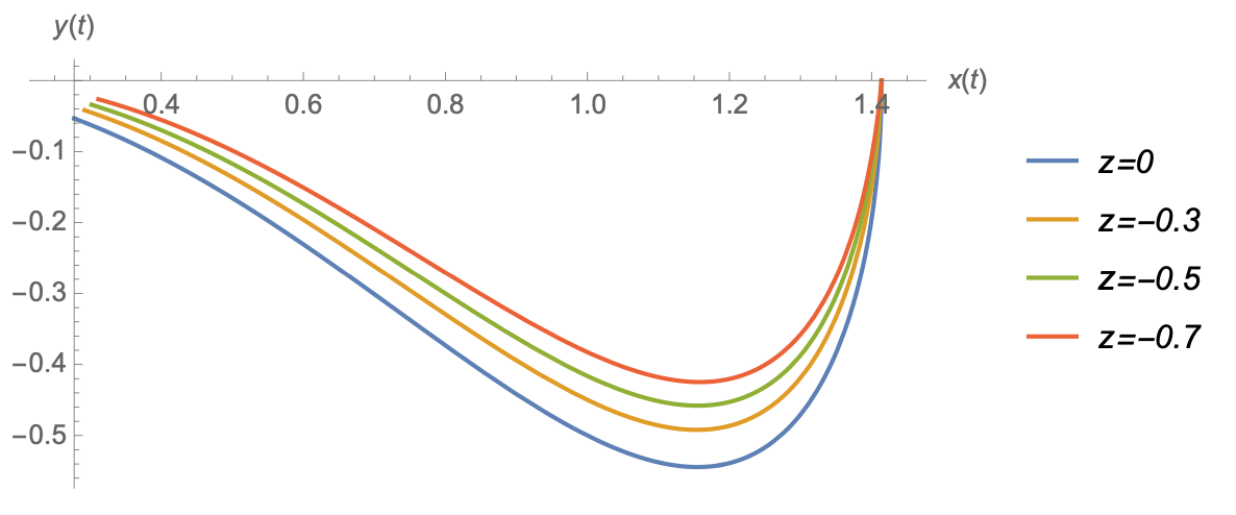}
\caption{\small First-order solutions (\ref{ee66}) of the equation (\ref{e6}) as in Figure~\ref{fig1} but with  negative values of the deformation parameter $z$.}
\centering
     \label{fig2}
\end{figure}

To illustrate the effect of the real deformation  parameter, we plot in Figure~\ref{fig1} several solutions (\ref{ee66}) for positive values of $z$  and in Figure~\ref{fig2} for the corresponding negative values.


  \subsection{Deformed case with $a(x)=1/x$ and arbitrary $ {b(t)} $}
  \label{s52}
  
  If $a(x)=x^{-1}$, then $ \xii(x)=x^{-1}$   $( x> 0)$ and from (\ref{d12}) we find that
 \be
\omega=\dfrac{ 1}{xy}\,\dd x\wedge \dd y ,\qquad h_{z,1}=\frac{y}{x}\, ,\qquad   
h_{z,2}=x\ln x\,\frac{\exp\bigl(-\frac{zy}{x}\bigr)-1}{zy}\, .
\nonumber
\ee
Proposition~\ref{prop2} leads to the corresponding  deformed system of Buchdahl equations for any $b(t)$:
\begin{equation} 
\begin{split}
\frac {\dd x}{\dd t}&  = y +  b(t)\,  x \ln x\left(  x \, \frac{\exp\bigl(\frac{zy}{x}\bigr)-1}{zy} -1\right)  \exp\biggl(-\frac{zy}{x}\biggr),\\[2pt]
\frac {\dd y}{\dd t} & =\frac {y^2}x + b(t)\left(  x \, \frac{1-\exp\bigl(-\frac{zy}{x}\bigr)}{z} \,(1+\ln x)  - y \exp\biggl(-\frac{zy}{x}\biggr)  \ln x\right)   
 .\end{split}
 \label{e9}
\end{equation}
 Its exact solution, provided by  Proposition~\ref{prop3}, reads as
\be
\begin{split}
\ln x(t) &=    \frac{\ln \left(1- z c_1 \eee^{\gam(t)} \right)}z   \left( \eee^{-\gam(t)} -z c_1   \right)     \left( c_2 +  \int^t    \frac{1}{\eee^{-\gam(\tau)}-z c_1 } \,\dd \tau \right), \\[4pt] 
y(t)&=x(t)\, \frac{\ln \left(1- z c_1 \eee^{\gam(t)} \right)}{z  } \,   .   \end{split}
\label{e10}
\ee
This yields a family of deformed Buchdahl-type systems, together with their exact solution, depending on the function $b(t)$. Once the latter is fixed, it is possible to study the first-order approximation in $z$, in analogy to the discussion in Section \ref{s51}.
 
Let us choose, for example,  $b(t)=1/t$ with $t\in  \mathbb R^\ast$  (so~$ \eee^{\gam(t)} =t$), as in Section~\ref{s32}. The exact solution (\ref{e10}) for the equations (\ref{e9}) reduces to
     \be
\begin{split}
\ln x(t) &=  \frac{ (1-z c_1 t)\ln\bigl(1-z c_1 t \bigr)}{zt}  \left( c_2  -  \frac{z c_1 t+\ln\bigl(1-z c_1 t \bigr)}{z^2 c^2_1} \right), \\[4pt] 
y(t)&=x(t)\, \frac{\ln \left(1- z c_1 t\right)}{z  } \,   .   \end{split}
\nonumber\nonumber
\ee
At the first-order in $z$, the system   (\ref{e9})  becomes
  \be
  \begin{split}
\frac {\dd x}{\dd t}&  \simeq y +  z\,  \frac{ y\ln x}{2t} \,  ,\\[2pt]
\frac {\dd y}{\dd t} & \simeq \frac {y^2}x +  \frac y t+ z\,  \frac{ y^2(\ln x-1)}{2tx} \,
 .\end{split}
 \label{e12}
\ee
After some algebraic manipulation, we arrive at the approximation of the deformed second-order  generalized Buchdahl equation (\ref{b01})  in the form
 \be
 \frac{\dd^2 x}{\dd t^2}\simeq\frac 1 x
 \left(\frac{\dd x}{\dd t}\right)^2+\frac 1t \,\frac{\dd x}{\dd t} - z\,\frac{\ln x}{2 t^2 }\,\frac{\dd x}{\dd t}  \, .
  \label{e13}
 \ee
As expected, the deformed equation only admits one Lie point symmetry 
$$
\displaystyle {\bf Y}= t\frac{\partial}{\partial t}+x\ln x\,\frac{\partial}{\partial x},
$$ 
and the  solution of (\ref{e12}) and (\ref{e13}) turns out to be 
\be
\begin{split}
x(t)&\simeq
\exp\left(- c_1c_2 -c_1  \frac{t^2}2\right) \left(1+z\,\frac {c_1^2 t}{2} \biggl( c_2-\frac {t^2}{6}  \biggl) \right) ,\\[2pt]
y(t) &\simeq -c_1 t \exp\left(- c_1c_2 -c_1  \frac{t^2}2\right) \left(1+z\,\frac {c_1 t}{2} \biggl(1+ c_1 c_2-\frac {c_1t^2}{6}  \biggl) \right),\end{split}
\nonumber
\ee
  that can be compared with the non-deformed equation (\ref{c20}).

 
  \subsection{Deformed case with $  a(x)=\aaa/x$  
  $(\aaa\ne 1)$ and arbitrary $b(t)$ }
    \label{s53}

As a last example, we consider the function $ \xii(x)=x^{-\aaa}$ $(x\in  \mathbb R^\ast)$ with $\aaa\ne 1$.
The expressions (\ref{d12}) yield
 \be
\omega=\dfrac{ 1}{x^\aaa y}\,\dd x\wedge \dd y ,\qquad h_{z,1}=\frac{y}{x^\aaa}\, ,\qquad h_{z,2}=x\, \frac{\exp\bigl(- {zy x^{-\aaa}} \bigr)-1}{zy(1-\aaa)} \,  , \qquad \aaa\ne 1,
\nonumber
\ee
in such a manner that the deformed system of Buchdahl equations (\ref{d14}) is now given by
\begin{equation} 
\begin{split}
\frac {\dd x}{\dd t}&  = y +  b(t)\,  \frac{x}{1-\aaa}  \left(  x^\aaa \,  \frac{\exp\bigl({zy x^{-\aaa}} \bigr)-1}{zy } -1\right)  \exp\bigl(- {zy x^{-\aaa}} \bigr),\\[2pt]
\frac {\dd y}{\dd t} & =\aaa\,\frac {y^2}x +  b(t)\,  \frac{1}{1-\aaa}  \left(  x^\aaa \,  \frac{\exp\bigl({zy x^{-\aaa}} \bigr)-1}{z } -\aaa y\right)  \exp\bigl(- {zy x^{-\aaa}} \bigr) .
\end{split}
 \label{e16}
\end{equation}
The corresponding exact solution, obtained by the previous prescription, turns out to be 
\be
\begin{split}
  x^{1-\aaa}(t) &=(1-\aaa) \,\frac{\ln \left(1- z c_1 \eee^{\gam(t)} \right)}z   \left( \eee^{-\gam(t)} -z c_1   \right)     \left( c_2 +  \int^t    \frac{1}{\eee^{-\gam(\tau)}-z c_1 } \,\dd \tau \right),\\[4pt] 
y(t)&=x^\aaa(t)\, \frac{\ln \left(1- z c_1 \eee^{\gam(t)} \right)}{z  } \,   .   \end{split}
\label{e17}
\ee
As in the preceding cases, we can obtain an approximation of the deformed system at the first-order in $z$, once an explicit function $b(t)$ has been chosen. To compare with the previous results, we again set $b(t)=1/t$ ($t\in  \mathbb R^\ast$). Under such an approximation,   the equations (\ref{e16}) read
  \be
  \begin{split}
\frac {\dd x}{\dd t}&  \simeq y +  z\,  \frac{ y\, x^{1-\aaa}}{2(1-\aaa)t} \,  ,\\[2pt]
\frac {\dd y}{\dd t} & \simeq \aaa\,\frac {y^2}x +  \frac y t- z\,  \frac{ y^2 x^{-\aaa}(1-2\aaa)}{2(1-\aaa)t}
\, ,\end{split}
 \label{e18}
\ee
which lead to the following deformed second-order Buchdahl equation   
 \be
 \frac{\dd^2 x}{\dd t^2}\simeq\frac \aaa x
 \left(\frac{\dd x}{\dd t}\right)^2+\frac 1t \,\frac{\dd x}{\dd t} + z\,\frac{x^{1-\aaa}}{2(\aaa-1) t^2 }\,\frac{\dd x}{\dd t}  \, .
  \label{e19}
 \ee
This ODE, again, only possesses one Lie point symmetry 
$$
\displaystyle {\bf Y}= (1-\alpha)t\frac{\partial}{\partial t}+x\frac{\partial}{\partial x},
$$
showing that it is not linearizable by point transformations \cite{Mah}. A solution of (\ref{e18}) and (\ref{e19}) can however be deduced from (\ref{e17}), obtaining that
\be
\begin{split}
x(t)&\simeq
\left( (\aaa-1)c_1 \biggl( c_2 +\frac{t^2}2\biggr) \right)^{\frac{1}{1-\aaa} } +\frac{z}{12}\,c_1^2 t(6c_2-t^2)
\left( (\aaa-1)c_1 \biggl( c_2 +\frac{t^2}2\biggr) \right)^{\frac{\aaa}{1-\aaa} } 
,\\[2pt]
y(t) &\simeq - \frac{z}{12}\,c_1^3 t^2 \bigl(6c_2(2\aaa-1)+t^2(2\aaa -3)\bigr)
\left( (\aaa-1)c_1 \biggl( c_2 +\frac{t^2}2\biggr) \right)^{\frac{2\aaa-1}{1-\aaa} }  \\[2pt]
&\qquad
- c_1 t\left( (\aaa-1)c_1 \biggl( c_2 +\frac{t^2}2\biggr) \right)^{\frac{\aaa}{1-\aaa} }.
\end{split}
\nonumber
\ee
Many other particular equations and their solution can be analyzed  applying the general results described in Propositions~\ref{prop2} and~\ref{prop3}.

 
  \section{Extending the generalized  Buchdahl equation  from the\\ oscillator algebra}
  \label{s6}
  
  So far we have obtained (deformed) generalized  Buchdahl equations from the 
 (quantum) book algebra $\mathfrak{b}_2$ in terms of arbitrary functions $a(x)$ and $b(t)$. Moreover, we have proven in Section~\ref{s35} that there is no possible extension of the generalized Buchdahl equation (\ref{b2}) from $\mathfrak{b}_2$, i.e.~from this underlying symmetry no additional $t$-dependent coefficient can be considered in a non-trivial way.
 
 Nevertheless, as $\mathfrak{b}_2$ arises as a subalgebra of other higher-dimensional Lie algebras, it is natural to extend the method of exact solutions to other LH systems (and their corresponding Poisson--Hopf deformations) that keep $\mathfrak{b}_2$ as a LH subalgebra.  From the classification of LH systems on the plane~\cite{LH2015,BHLS}, it follows  that the relevant candidates are the oscillator $\mathfrak{h}_4$,  $\mathfrak{sl}(2,\mathbb{R})$ and the so-called two-photon $\mathfrak{h}_6$ LH algebras.
 
Taking into account the canonical representation (\ref{c5}), we find that the simplest extension is provided by  the oscillator LH algebra, corresponding to  the imprimitive class ${\rm I}_{8}$ in~\cite{LH2015,BHLS}, that entails the introduction of an additional non-trivial arbitrary $t$-dependent coefficient. Furthermore, the corresponding exact solution is straightforward, allowing us to apply this result to the framework of Buchdahl equations, that we develop in what follows. We recall that the embedding $\mathfrak{b}_2\subset \mathfrak{h}_4 $   has already been used in the context of $t$-dependent epidemic models in~\cite{CFH}. However,  we   stress that an exact solution can also be deduced for the deformed counterpart  from a quantum oscillator algebra, which, to  best of our knowledge, was still lacking, and will be addressed to in the next section.

 Thus, we start with the oscillator $\mathfrak{h}_4$-LH algebra with Hamiltonian functions expressed in   canonical coordinates $(q,p)$ by~\cite{LH2015,BHLS}
 \be
 h_1=  - q , \qquad h_2=  qp,\qquad h_3=  p,\qquad h_0=  1,
\label{g1}
\ee
obeying the commutation relations
 \be
\{ h_2,h_1\}_\wcan= - h_1  ,\qquad\{ h_2,h_3\}_\wcan=   h_3  ,\qquad\{ h_3,h_1\}_\wcan= h_0  ,
\qquad \{ h_0,\,\cdot\,\}_\wcan=0 ,
\label{g2}
\ee
with respect to the canonical symplectic form  (\ref{c4}). Hence, $h_2$ can be regarded as the number 
 generator, $h_1$, $h_3$ as lowering/raising generators and $h_0$ as the central element (necessary to close the brackets). The extension of the $t$-dependent Hamiltonian (\ref{c6}) in terms of two arbitrary $t$-dependent functions $b_1(t)\equiv b(t)$ and $b_2(t)$ yields
\begin{equation} 
 h_t= h_1 + b_1(t)  h_2  + b_2(t)  h_3     = - q+  b_1(t)  qp +  b_2(t)  p,  
 \label{g3}
\end{equation}
giving rise to the Hamilton equations
  \begin{equation} 
\frac{\dd q}{\dd t}=\{q, h_t\}_\wcan=b_1(t) q +b_2(t),
\qquad
\frac{\dd p}{\dd t}=\{p, h_t\}_\wcan=1-b_1(t) p  ,
 \label{g33}
\end{equation}
which form again a linear and uncoupled system. 

The associated Hamiltonian vector fields are obtained from (\ref{g1}) through the relation (\ref{b6}) and read as
\be
 \>X_1=\frac{\partial}{\partial p}\, , 
 \qquad     \>X_2=q\, \frac{\partial}{\partial q}- p\, \frac{\partial}{\partial p}\, ,   
 \qquad     \>X_3= \frac{\partial}{\partial q}\, ,
 \label{g5}    
\ee
which satisfy the Lie commutators
\be
[\>X_2,\>X_1]=\>X_1,\qquad [\>X_2,\>X_3]=-\>X_3,\qquad [\>X_1,\>X_3]= 0.
 \label{g55}    
\ee
Therefore, the Vessiot--Guldberg Lie algebra is isomorphic to the  (1+1)-dimensional Poincar\'e algebra in a light-cone basis with $\>X_2$ playing the role of  the boost generator and  $\>X_1$, $\>X_3$ as translations. In fact, if $(q,p)$ are identified with the  light-like coordinates $(x_+,x_-)$, the vector fields (\ref{g5}) are just the Killing vector fields of the metric $\dd s^2=\dd x_+\dd x_- $ in the Minkowskian spacetime. 

The oscillator LH system (\ref{g33}) can then be solved directly leading to the following exact solution in terms of 
 two integration constants $c_1$ and $c_2$:
  \be
\begin{split}
q(t)&= \left( c_1 +  \int^t \eee^{-\gam(\tau )}b_2(\tau) \dd \tau\right)  \eee^{\gam(t)} ,\qquad \gam(t):= \int^t b_1(\tau)\dd \tau ,\\[4pt]
p(t)&=\left( c_2 +  \int^t \eee^{\gam(\tau )}  \dd \tau\right)  \eee^{-\gam(t)}.
\end{split}
\label{g6}
\ee

The above results can be applied in a straightforward manner to the generalized Buchdahl equations by applying 
 the change of coordinates (\ref{c3}). The Hamiltonian functions (\ref{g1}) and vector fields (\ref{g5})
 turn out to be
 \be
 h_1= y \, \Xi(x), \qquad h_2=  -\int^x \Xi(\xi)\dd \xi,\qquad h_3=  \frac{1}{y\,\Xi(x) }\int^x \Xi(\xi)\dd \xi,\qquad h_0=  1,
\label{g77}
\ee
\begin{equation}
\begin{split}
& \>X_1=y\frac{\partial }{\partial x}+a(x)y^2\frac{\partial}{\partial y}\, ,\qquad  \>X_{2}=y\dfrac{\partial}{\partial y}\, ,\\[2pt]   
&\>X_3=-   \frac{1}{y\,\Xi^2(x) }\int^x \Xi(\xi)\dd \xi \,\frac{\partial }{\partial x}-\frac{1}{\Xi^2(x) }\left(\Xi(x)  +a(x)\int^x \Xi(\xi)\dd \xi \right)\frac{\partial}{\partial y} \, ,
 \label{g8}
 \end{split}
 \end{equation}
with $\Xi(x)$ given in (\ref{c10}). They fulfill the same commutation rules (\ref{g2}) and (\ref{g55}) (the former with respect to the symplectic form (\ref{b7})). Observe that, as a byproduct, the $\>X_i$ in (\ref{g8}) are 
Killing vector fields of the Minkowskian metric, now  reading as
\begin{equation}
\begin{split}
&\dd s^2=a(x)\left(\Xi(x)+a(x)\int^x \Xi(\xi)\dd \xi  \right)\dd x^2+\frac 1{y^2}\int^x \Xi(\xi)\dd \xi  \,\dd y^2\\[2pt]
&\qquad \qquad -\frac 1 y \left(\Xi(x) + 2 a(x) \int^x \Xi(\xi)\dd \xi \right)\dd x\,\dd y.
\end{split}
\nonumber
 \end{equation}

From $h_t$ (\ref{g3}) or $\>X_t= \>X_1 + b_1(t)  \>X_2  + b_2(t)  \>X_3$, we  arrive at the extended generalized Buchdahl equations as  the first-order nonlinear system of ODEs given by
 \begin{equation}
 \begin{split}
&  \frac{\dd x}{\dd t}=y - b_2(t) \,\frac{1}{y\,\Xi^2(x) }\int^x \Xi(\xi)\dd \xi \,  ,\\
& \frac{\dd y}{\dd t}=a(x)y^2+b_1(t)y -b_2(t)\,\frac{1}{\Xi^2(x) }\left(\Xi(x)  +a(x)\int^x \Xi(\xi)\dd \xi \right)  ,
 \end{split}
 \label{g9}
 \end{equation}
(compare with the system (\ref{b2})). In this respect, it should be observed that the consideration of the 
$\mathfrak{h}_4$-LH algebra implies the introduction of a `naive' term $b_2(t)p$ in the Hamiltonian $h_t$ (\ref{g3}), whose effect becomes quite strong in the context of the Buchdahl equations above.
In particular, the coefficient $b_2(t)$ introduces complicated terms in both equations with   $\frac{\dd x}{\dd t}\ne y$.
Their corresponding exact solution is provided by (\ref{g6}) and characterized by the  following `extended' version of Proposition~\ref{prop1}.

  \begin{proposition}
 \label{prop4}
The general solution of the extended generalized Buchdahl equations (\ref{g9}), determined by the oscillator $\mathfrak{h}_4$-LH algebra, is given by
  \be
\begin{split}
&\int^x \xii(\xi)\dd \xi = - \left( c_1 +  \int^t \eee^{-\gam(\tau )}b_2(\tau) \dd \tau\right) \left( c_2 +  \int^t \eee^{\gam(\tau)}  \dd \tau\right)  ,  \\[4pt]
& y(t)= - \frac{ \eee^{\gam(t)}}{\Xi(x) }\left( c_1 +  \int^t \eee^{-\gam(\tau )}b_2(\tau) \dd \tau\right)  ,\\[4pt]
&   \xii(x):=\exp\left(-\int^x a(\xi)\dd \xi\right),\qquad \gam(t):= \int^t b_1(\tau)\dd \tau ,
 \end{split}
\label{g11}
\ee
where $c_1$ and $c_2$ are the two  constants of integration determined by the  initial conditions.
 \end{proposition}

 Once the coefficients $b_1(t)$, $b_2(t)$ and $a(x)$  have been chosen, the procedure to obtain the solution $(x(t),y(t))$ from Proposition~\ref{prop4} consists in calculating $\xii(x)$ and $\gam(t)$, introduce them in the first equation trying to compute $x(t)$,  and substituting it into the second equation, finding $y(t)$. 
 
 We point out that the particular case with constant $b_2(t)= b_0$, covered by Proposition~\ref{prop4}, is also endowed with the    $\mathfrak{h}_4$-LH algebra. Although $ h_t$ (\ref{g3}) is thus composed of only two Hamiltonian functions, a third function is needed to close the Poisson brackets leading to a LH algebra isomorphic to $\mathfrak{h}_4$. Explicitly, if we define $h_\pm := h_1\pm b_0 h_3$, then the commutation rules  
  (\ref{g2}) are transformed into 
 \be
\{ h_2,h_\pm\}_\omega= - h_\mp     ,\qquad\{ h_+,h_-\}_\omega=2 b_0 h_0  ,
\qquad \{ h_0,\,\cdot\,\}_\omega=0 ,
\nonumber
\ee
with respect to the symplectic form (\ref{b7}), and $h_t= h_+ +b_1(t)h_2$. Likewise,  this particular system comes from the $t$-dependent vector field $\>X_t= \>X_+ + b_1(t)  \>X_2 $ where $\>X_\pm: = \>X_1\pm b_0 \>X_3$ such that 
$  [\>X_2,\>X_\pm]= \>X_\mp$ and $ [\>X_+,\>X_-]= 0$, being isomorphic to the Poincar\'e Lie algebra.
   
  We illustrate the   $\mathfrak{h}_4$-LH systems (\ref{g9}) together with their general solution (\ref{g11}) by constructing the extensions of the particular generalized Buchdahl equations studied previously in Sections~\ref{s31}--\ref{s33}, the final results of which are summarized in Table~\ref{table1}.

   \begin{table}[h!]
{\small
\caption{ \small{Extensions of the particular generalized  Buchdahl equations of Sections~\ref{s31}--\ref{s33}  from the oscillator LH algebra. For each case, we indicate: the choice of the coefficients  $a(x)$ and $b_1(t)$ for arbitrary $b_2(t)$, the symplectic form (\ref{b7}), the functions $\Xi(x)$ and $\gam(t)$, the Hamiltonian functions (\ref{g77}), the system (\ref{g9}) and its corresponding general solution  (\ref{g11}).}}
  \begin{center}
\noindent 
\begin{tabular}{ l}
\hline

\hline
\\[-6pt]
$\bullet$ {\bf  Case I}\quad Extended Buchdahl equation\quad $a(x)=    3 x^{-1}$\quad $b_1(t)=     t^{-1}$\quad $  \xii(x)=   {x^{-3}} $\quad $\gam(t )=\ln t $\\[4pt]
   $\displaystyle{\quad  \omega=\dfrac{ 1}{x^3y}\,\dd x\wedge \dd y \qquad h_1=\frac{y}{x^3} \qquad h_2=\frac{1}{2x^2}\qquad h_3=-\frac{x}{2y} }$
   \\[8pt]
 $\displaystyle{\quad   \frac{\dd x}{\dd t}=y + b_2(t)\frac{x^4}{2y} \qquad   \frac{\dd y}{\dd t}=\frac{3y^2}{x} +\frac{y}{t}+b_2(t) \frac{x^3}{2} }$
    \\[8pt]
    $\displaystyle{\quad   x(t)=\pm\left\{ \left(c_1+\int^t\frac{b_2(\tau)}{\tau}\,\dd\tau \right)\left(2 c_2+t^2 \right)    \right\}^{-1/2} }$\\[8pt]
         $\displaystyle{\quad    y(t)=\mp\, t \left(c_1+\int^t\frac{b_2(\tau)}{\tau}\,\dd\tau \right)^{-1/2} \left(2 c_2+t^2 \right)^{-3/2} }$
    \\[14pt]

\hline
\\[-6pt]
$\bullet$  {\bf  Case II} \quad  $a(x)=      x^{-1}$\quad arbitrary $b_1(t)$  \quad $ \xii(x)=\ {x^{-1}}$
     \\[4pt]
 $\displaystyle{\quad  \omega=\dfrac{ 1}{x y}\,\dd x\wedge \dd y \qquad h_1=\frac{y}{x } \qquad h_2=- \ln x\qquad h_3= \frac{x\ln x}{y} }$
   \\[8pt]
 $\displaystyle{\quad   \frac{\dd x}{\dd t}=y - b_2(t)\frac{x^2\ln x}{y} \qquad   \frac{\dd y}{\dd t}=\frac{y^2}{x} +b_1(t) {y} -b_2(t) x \bigl( 1+\ln x\bigr)  }$
    \\[8pt]
    $\displaystyle{\quad   x(t)= \exp\left\{ -  \left(c_1+\int^t \eee^{-\gam(\tau )}{b_2(\tau)}\dd\tau \right)     \left(  c_2+\int^t \eee^{\gam(\tau )}\dd\tau \right)  \right\}  }$\\[8pt]
         $\displaystyle{\quad    y(t)=-\eee^{\gam(t)}\left(c_1+\int^t \eee^{-\gam(\tau )}{b_2(\tau)}\dd\tau \right) \exp\left\{ -  \left(c_1+\int^t \eee^{-\gam(\tau )}{b_2(\tau)}\dd\tau \right)     \left(  c_2+\int^t \eee^{\gam(\tau )}\dd\tau \right)  \right\}  }$
    \\[12pt]
     
\quad  {\bf  Subcase II} \quad  $a(x)=      x^{-1}$\quad   $b_1(t)=t^{-1}$  \quad $ \xii(x)=\ {x^{-1}}$
\quad $\gam(t )=\ln t $      \\[6pt]
     $\displaystyle{\quad   x(t)= \exp\left\{ -  \left(c_1+\int^t  \frac{b_2(\tau)}{\tau}\,\dd\tau \right)     \left(  c_2+ \frac{t^2}2  \right)  \right\}  }$\\[10pt]
         $\displaystyle{\quad    y(t)=- t\left(c_1+\int^t  \frac{b_2(\tau)}{\tau}\,\dd\tau \right) \exp\left\{ -  \left(c_1+\int^t  \frac{b_2(\tau)}{\tau}\,\dd\tau\right)     \left(  c_2+ \frac{t^2}2\right)  \right\}  }$
\\[14pt]
     
 \hline
\\[-6pt]
$\bullet$  {\bf  Case III} \quad $a(x)=    \alpha\,  x^{-1}\ (\alpha\ne 1)$\quad arbitrary $b_1(t)$  \quad $ \xii(x)=\ {x^{-\alpha}}$
     \\[4pt]
 $\displaystyle{\quad  \omega=\dfrac{ 1}{x^\alpha y}\,\dd x\wedge \dd y \qquad h_1=\frac{y}{x^\alpha } \qquad h_2=- \frac{x^{1-\alpha}}{1-\alpha}  \qquad h_3= \frac{x }{y(1-\alpha)} }$
   \\[8pt]
 $\displaystyle{\quad   \frac{\dd x}{\dd t}=y - b_2(t)\frac{x^{1+\alpha}}{(1-\alpha)y} \qquad   \frac{\dd y}{\dd t}=\frac{\alpha\, y^2}{x} +b_1(t) {y} -b_2(t) \frac{x^\alpha}{1-\alpha}    }$
    \\[8pt]
    $\displaystyle{\quad   x(t)=   \left\{ (\aaa-1) \left(c_1+\int^t \eee^{-\gam(\tau )}{b_2(\tau)}\dd\tau \right) \left( c_2 +  \int^t \eee^{\gam(\tau)}  \dd \tau \right) \right\}^{\frac{1}{1-\aaa} }   }$\\[8pt]
         $\displaystyle{\quad    y(t)=-\eee^{\gam(t)}\left(c_1+\int^t \eee^{-\gam(\tau )}{b_2(\tau)}\dd\tau \right)  \left\{ (\aaa-1) \left(c_1+\int^t \eee^{-\gam(\tau )}{b_2(\tau)}\dd\tau \right) \left( c_2 +  \int^t \eee^{\gam(\tau)}  \dd \tau \right) \right\}^{\frac{\aaa}{1-\aaa} }   }$\\[12pt]
     
\quad  {\bf  Subcase III} \quad  $a(x)=    \alpha\,  x^{-1}\ (\alpha\ne 1)$\quad   $b_1(t)=t^{-1}$  \quad $ \xii(x)=\ {x^{-\alpha}}$
\quad $\gam(t )=\ln t $        \\[6pt]
    $\displaystyle{\quad   x(t)=   \left\{ (\aaa-1) \left(c_1+\int^t  \frac{b_2(\tau)}{\tau}\,\dd\tau \right) \left( c_2 +   \frac{t^2}2 \right) \right\}^{\frac{1}{1-\aaa} }   }$\\[8pt]
         $\displaystyle{\quad    y(t)=- t\left(c_1+\int^t  \frac{b_2(\tau)}{\tau}\,\dd\tau\right)  \left\{ (\aaa-1) \left(c_1+\int^t  \frac{b_2(\tau)}{\tau}\,\dd\tau \right) \left( c_2 + \frac{t^2}2\right) \right\}^{\frac{\aaa}{1-\aaa} }   }$    \\[14pt]
\hline

\hline
\end{tabular}
 \end{center}
\label{table1}
}
\end{table}


\section{Deformed generalized Buchdahl  equation from the quantum\\   oscillator algebra}
  \label{s7}

  Among all possible quantum deformations of the oscillator Lie algebra $\mathfrak{h}_4$~\cite{h4}, only the so-called non-standard deformation, $U_{z}(\mathfrak{h}_4)\equiv \mathfrak{h}_{z,4 }$, enables the Hopf algebra embedding $\mathfrak{b}_{z,2}\subset \mathfrak{h}_{z,4 }$. In fact, $\mathfrak{h}_{z,4 }$ is a central extension of the non-standard  quantum (1+1) Poincar\'e algebra~\cite{poincare2,boson} in the light-cone basis (\ref{g55}). In what follows, we first recall the role of   $\mathfrak{h}_{z,4 }$ in the framework of LH systems and then apply these results to the context of Buchdahl equations, arriving at the extension of the  deformed  generalized equations obtained in Sections~\ref{s4} and \ref{s5}.


\subsection{Quantum  oscillator algebra and deformed oscillator LH systems}
  \label{s71}
  
  The Hopf structure of $\mathfrak{h}_{z,4 }$ is determined by the following coproduct map 
  and   compatible commutation relations in a  basis $(v_1,v_2,v_3,v_0)$~\cite{h4}:
 \be
\begin{split}
\Delta_z(v_1)&=v_1\otimes 1 + 1 \otimes v_1 , \qquad  \Delta_z(v_2) =v_2\otimes \eee^{-z v_1}+1\otimes v_2, \\[2pt]
\Delta_z(v_3)&=v_3\otimes \eee^{-z v_1}+1\otimes v_3 + z  v_2 \otimes \eee^{-z v_1}   v_0 
 ,  \qquad \Delta_z(v_0)=v_0\otimes 1 + 1 \otimes v_0 , \\[1pt]
[v_2,v_1]_z&=  \frac{\eee^{-z v_1}-1}{z} \, , \qquad [v_2,v_3]_z = v_3 ,\qquad [v_3,v_1]_z = \eee^{-z v_1}v_0,\qquad [v_0,\,\cdot\,]_z = 0.
\end{split}
\label{h1}
\ee
 In canonical variables $(q,p)$, a deformed  symplectic representation $D_z$ of $\mathfrak{h}_{z,4 }$  turns out to be~\cite{boson}
  \be
  \begin{split}
 h_{z,1}:&= D_z(v_1)=  -q , \qquad\ \,  h_{z,2}:= D_z(v_2)=  \left( \frac{ \eee^{z q} -1}{z}\right) p,\\[2pt] 
 h_{z,3}:&= D_z(v_3)=   \eee^{z q} p,\qquad h_{z,0}:= D_z(v_0)=   1,
\label{h3}
\end{split}
\ee
which fulfill the deformed Poisson brackets  given by
\be
  \begin{split}
\{h_{z,2},h_{z,1}\}_\wcan&=\frac{\eee^{-z h_{z,1}} -1}{z}\,,\qquad \{h_{z,2},h_{z,3}\}_\wcan=h_{z,3},\\[2pt]
  \{h_{z,3},h_{z,1}\}_\wcan&=\eee^{-z h_{z,1}}h_{z,0},\qquad\ \, \{h_{z,0},\,\cdot\,\}_\wcan=0,
\label{h4}
\end{split}
\ee
with respect to canonical symplectic form  (\ref{c4}).

The relation $\iota_{{\bf X}_{z,i}}\wcan={\rm d}h_{z,i}$ leads to the associated deformed  Hamiltonian vector fields  
\be
 \>X_{z,1}= \frac{\partial}{\partial p} \, ,   \qquad  \>X_{z,2}=\left( \frac{ \eee^{z q}-1}{z} \right)\frac{\partial}{\partial q}- \eee^{z q} p \, \frac{\partial}{\partial p} \, ,\qquad \>X_{z,3}=\eee^{z q} \frac{\partial}{\partial q}  -z  \eee^{z q} p \,  \frac{\partial}{\partial p}    \, , 
\nonumber
\ee
 which do not close on the non-standard quantum Poincar\'e algebra~\cite{poincare2}, but on 
   a smooth  Stefan--Sussmann distribution~\cite{Vaisman,Pa57,WA} given by the commutation relations~\cite{Ballesteros6}
 \be
 [{\bf X}_{z,2},{\bf X}_{z,1}]=  \eee^{-z h_{z,1}} \, {\bf X}_{z,1},\qquad [{\bf X}_{z,2},{\bf X}_{z,3}]= -  {\bf X}_{z,3}, \qquad [{\bf X}_{z,3},{\bf X}_{z,1}]= z\, \eee^{-z h_{z,1}}h_{z,0} \, {\bf X}_{z,1}.
 \label{h6}
 \ee
 Thus, we obtain a deformed $t$-dependent Hamiltonian and  vector field  in terms of two coefficients $b_1(t)\equiv b(t)$ and $b_2(t)$ as
\begin{equation} 
\begin{split}
h_{z,t}&=h_{z,1} + b_1(t)  h_{z,2}   + b_2(t)  h_{z,3}   =  -  q  + b_1(t)   \biggl( \frac{ \eee^{z q} -1}{z}\biggr) p   + b_2(t)  \eee^{z q} p, \\ 
  \>X_{z,t}&=\>X_{z,1}  + b_1(t)   \>X_{z,2}  + b_2(t)   \>X_{z,3}  \\[2pt]
  & =  \frac{\partial}{\partial p}+ b_1(t) \left( \left( \frac{ \eee^{z q}-1}{z} \right)\frac{\partial}{\partial q}- \eee^{z q}p \, \frac{\partial}{\partial p} \right) +b_2(t) \left(  \eee^{z q} \frac{\partial}{\partial q}  -z  \eee^{z q} p \,  \frac{\partial}{\partial p}  \right)     ,
\end{split}
\label{hh66}
\end{equation}
yielding  the first-order system of  nonlinear and coupled  ODEs on $\mathbb R^2$  given by
\begin{equation} 
\begin{split}
\frac{\dd q}{\dd t}&=  b_1(t) \biggl( \frac{ \eee^{z q}-1}{z} \biggr)  +b_2(t)  \eee^{z q }, \\[2pt]
\frac{\dd p}{\dd t}&=1 - b_1(t) \eee^{z q} p -  z b_2(t)  \eee^{z q} p . 
\end{split}
 \label{h8}
\end{equation}
The first  equation can be  solved directly and by substituting into the second one, we obtain the general solution  
\be
\begin{split}
q(t)&=-\frac 1 z \, {\ln \left\{ 1- z \eee^{\gam(t)} \left(c_1 + \int^t \eee^{-\gam(\tau)}b_2(\tau) \,\dd \tau\right)\right\}}    ,\qquad \gam(t):= \int^t b_1(\tau)\dd \tau ,\\[4pt]
 p(t)&=   \left\{ \eee^{-\gam(t)} -z \left(c_1 + \int^t \eee^{-\gam(\tau)}b_2(\tau) \,\dd \tau\right)  \right\}\\[4pt]
 &\qquad \times \left\{ c_2 +  \int^t   \left(\eee^{-\gam(\tau)}-z   \left(c_1 + \int^\tau \eee^{-\gam(\tau')}b_2(\tau') \,\dd \tau'\right)  \right)^{-1}  \dd \tau \right\}  ,
\end{split}
\label{h9}
\ee
where  $c_1$ and $c_2$ are the two  integration constants.  
We remark that, to best of our knowledge, 
this exact solution has not yet been considered in the framework of LH systems. On the contrary, the usual approach is to deduce `deformed' superposition rules from $t$-independent constants of the motion~\cite{Ballesteros6}. Note also that,  as expected, the expressions (\ref{g1})--(\ref{g6}) are recovered from (\ref{h3})--(\ref{h9}) under the limit $z\to 0$.
 
The first-order  of the  power series expansion in $z$ of  (\ref{h8})  yields
 \begin{equation} 
\begin{split}
\frac{\dd q}{\dd t}&= b_2(t) + b_1(t)q+z\biggl( b_2(t)  q  +\frac 12 b_1(t)   q^2 \biggr) +o[z^2]   , \\[2pt]
\frac{\dd p}{\dd t}&= 1 - b_1(t) p-z \bigl(   b_2(t)   p +  b_1(t)   qp   \bigr) +o[z^2]   ,
\end{split}
\nonumber
\end{equation}
to be compared with  (\ref{d11}).


 \subsection{Deformed extended generalized Buchdahl  equation and its general solution}
  \label{s72}

Similarly to Section~\ref{s42}, we apply the  change of variables (\ref{c3}) to the expressions (\ref{h3}) and (\ref{h8})   obtaining  the 
deformation of the  extended generalized Buchdahl equation of Section~\ref{s6}  from the quantum oscillator algebra $\mathfrak{h}_{z,4 }$.

  \begin{proposition}
 \label{prop5}
(i)  The deformation of the $t$-dependent extended generalized Buchdahl Hamiltonian (\ref{g3})  is defined,   in terms of the variables $(x,y)$, by the Hamiltonian functions 
\begin{equation} 
\begin{split}
h_{z,t}&=h_{z,1} + b_1(t)  h_{z,2}  + b_2(t)  h_{z,3}   , \\[2pt]
h_{z,1} & =y\, \xii(x)   ,\qquad   h_{z,2}   =\frac{\exp\bigl(-z y\, \xii(x)\bigr) -1}{z y\, \xii(x)}\int^x \xii(\xi)\,\dd \xi  \, , \\[2pt] 
h_{z,3} & =\frac{\exp\bigl(-z y\, \xii(x)\bigr) }{y\, \xii(x)}\int^x \xii(\xi)\,\dd \xi  \, , \qquad
\xii(x):=\exp\left(-\int^x a(\xi)\dd \xi\right) , 
\end{split}
 \label{h12}
\end{equation}
which together with $h_{z,0}=1$ verify the same deformed Poisson brackets (\ref{h4}), now with respect to the (non-canonical) symplectic form $\omega$ (\ref{c2}).

\noindent
(ii) The  Poisson--Hopf deformation of the  extended generalized Buchdahl system (\ref{g9}) reads as
\begin{equation} 
\begin{split}
\frac {\dd x}{\dd t}&=\{x, h_{z,t}\}_\omega = y + b_1(t) \left( \frac{\exp\bigl(z y\, \xii(x)\bigr) -1 -z y\,\xii(x) }{z y\,\xii^2(x)}\right)  \exp\bigl(-z y\, \xii(x)\bigr) \int^x \xii(\xi)\,\dd \xi   \\[2pt]
&\qquad\qquad\qquad\quad  -  b_2(t) \left( \frac{  1+ z y\, \xii(x)}{  y\,\xii^2(x)}\right)  \exp\bigl(-z y\, \xii(x)\bigr) \int^x \xii(\xi)\,\dd \xi  \, , \\[2pt]
\frac {\dd y}{\dd t}&=\{y, h_{z,t}\}_\omega = a(x) y^2 + b_1(t)  \exp\bigl(-z y\, \xii(x)\bigr) \\[2pt]
&\qquad\qquad  \times \left(     \frac{\exp\bigl(z y\, \xii(x)\bigr) -1}{z \, \xii(x)} + \frac{ \exp\bigl(z y\, \xii(x)\bigr) -1 -z y\,\xii(x) }{z \,\xii^2(x)}\, a(x) \int^x \xii(\xi)\,\dd \xi
  \right)  \\[2pt]
&\qquad\qquad\qquad\quad  - b_2(t) \, \frac{\exp\bigl(-z y\, \xii(x)\bigr)  }{ \xii^2(x)} \left( \xii(x)+\bigl(1+ z y \, \xii(\xi)\bigr) a(x) \int^x \xii(\xi)\,\dd \xi\right)
    ,\end{split}
 \label{h13}
\end{equation}
for arbitrary $a(x)$, $b_1(t)$, $b_2(t)$  and $z$.
 \end{proposition}

The  deformed vector fields associated to the Hamiltonian functions (\ref{h12}) turn out to be 
 \begin{equation} {\small
\begin{split}
 \>X_{z,t}&=\>X_{z,1}+b_1(t)\>X_{z,2} +b_2(t)\>X_{z,3}  ,\qquad 
 \>X_{z,1}=y\frac{\partial }{\partial x}+a(x)y^2\frac{\partial}{\partial y}\, ,\\[2pt]
 \>X_{z,2}&=\left( \frac{\exp\bigl(z y\, \xii(x)\bigr) -1 -z y\,\xii(x) }{z y\,\xii^2(x)}\right)  \exp\bigl(-z y\, \xii(x)\bigr) \int^x \xii(\xi)\,\dd \xi \, \dfrac{\partial}{\partial x}\\[2pt]
 &\qquad +    \exp\bigl(-z y\, \xii(x)\bigr)  \left(     \frac{\exp\bigl(z y\, \xii(x)\bigr) -1}{z \, \xii(x)} + \frac{ \exp\bigl(z y\, \xii(x)\bigr) -1 -z y\,\xii(x) }{z \,\xii^2(x)}\, a(x) \int^x \xii(\xi)\,\dd \xi
  \right)\dfrac{\partial}{\partial y}\,   ,\\[2pt]
  \>X_{z,3}&=-\left( \frac{  1+ z y\, \xii(x)}{  y\,\xii^2(x)}\right)  \exp\bigl(-z y\, \xii(x)\bigr) \int^x \xii(\xi)\,\dd \xi   \, \dfrac{\partial}{\partial x}\\[2pt]
  &\qquad - \frac{\exp\bigl(-z y\, \xii(x)\bigr)  }{ \xii^2(x)} \left( \xii(x)+\bigl(1+ z y \, \xii(\xi)\bigr) a(x) \int^x \xii(\xi)\,\dd \xi\right),
   \end{split}
\nonumber\nonumber }
\end{equation}
  that  span a distribution with the same commutation rules given by (\ref{h6}).

From the general solution of the deformed extended generalized Buchdahl equations  in canonical variables (\ref{h9}), we arrive at the one corresponding to the deformed system in Proposition~\ref{prop5}.

 \begin{proposition}
  \label{prop6}
The general solution of the  first-order system of deformed extended generalized Buchdahl equations (\ref{h13})     is given by
  \be
\begin{split}
\int^x \xii(\xi)\dd \xi &= \frac 1 z \, {\ln \left\{ 1- z \eee^{\gam(t)} \left(c_1 + \int^t \eee^{-\gam(\tau)}b_2(\tau) \,\dd \tau\right)\right\}}  \left\{ \eee^{-\gam(t)} -z \left(c_1 + \int^t \eee^{-\gam(\tau)}b_2(\tau) \,\dd \tau\right)  \right\}  \\[4pt] 
& \qquad\qquad \times \left\{ c_2 +  \int^t   \left(\eee^{-\gam(\tau)}-z   \left(c_1 + \int^\tau \eee^{-\gam(\tau')}b_2(\tau') \,\dd \tau'\right)  \right)^{-1}  \dd \tau \right\} , \\[4pt] 
y(t)&=   \frac 1{z\,  \xii(x)}\, {\ln \left\{ 1- z \eee^{\gam(t)} \left(c_1 + \int^t \eee^{-\gam(\tau)}b_2(\tau) \,\dd \tau\right)\right\}} ,
\\[4pt] 
  \xii(x)&:=\exp\left(-\int^x a(\xi)\dd \xi\right) ,\qquad  \gam(t):= \int^t b(\tau)\dd \tau\, ,
 \end{split}
\label{h14}
\ee
where $c_1$ and $c_2$ are the two  integration constants  provided by the  initial conditions.
 \end{proposition}

   \begin{table}[t!]
{\footnotesize
\caption{ \small{Extensions of the particular deformed generalized  Buchdahl equations of Sections~\ref{s51}--\ref{s53}  from the deformed oscillator LH algebra. For each case we show the choice of the coefficients  $a(x)$ and $b_1(t)$, always with arbitrary $b_2(t)$, the symplectic form (\ref{b7}), the functions $\Xi(x)$ and $\gam(t)$, the Hamiltonian functions (\ref{h12}), the system (\ref{h13})  and its corresponding general solution  (\ref{h14}).}}
  \begin{center}
\noindent 
\begin{tabular}{ l}
\hline

\hline
\\[-6pt]
$\bullet$ {\bf  Case I}\quad Deformed extended Buchdahl equation\quad $a(x)=    3 x^{-1}$\quad $b_1(t)=     t^{-1}$\quad $  \xii(x)=   {x^{-3}} $\quad $\gam(t )=\ln t $\\[3pt]
   $\displaystyle{\quad  \omega=\dfrac{ 1}{x^3y}\,\dd x\wedge \dd y \qquad h_{z,1}=\frac{y}{x^3} \qquad h_{z,2}=x\,\frac{1-\exp\bigl(-\frac{zy}{x^3}\bigr)}{2zy} \qquad h_{z,3}=-\frac{x}{2y} \exp\biggl(-\frac{zy}{x^3}\biggr)}$
   \\[7pt]
 $\displaystyle{\quad  \frac {\dd x}{\dd t}   = y + \frac 1{2t}  \left( x \exp\biggl(-\frac{zy}{x^3}\biggr)- x^4\, \frac{1-\exp\bigl(-\frac{zy}{x^3}\bigr)}{zy} \right)  +b_2(t)\,\frac{x^4+z x y }{2 y} \exp\biggl(-\frac{zy}{x^3}\biggr) }$   \\[12pt]
  $\displaystyle{\quad 
  \frac {\dd y}{\dd t}   =\frac {3y^2}x   +\frac 1{2 t}\left( 3 y \exp\biggl(-\frac{zy}{x^3}\biggr)- x^3\,  \frac{1-\exp\bigl(-\frac{zy}{x^3}\bigr)}{z}\right) +\frac 12 b_2(t) \bigl(x^3+ 3 z y \bigr)  \exp\biggl(-\frac{zy}{x^3}\biggr)}$
    \\[12pt]
    $\displaystyle{\quad    \frac{1}{x^2(t)} = \frac{2}{zt} \left(z t  \left(c_1 + \int^t \frac {b_2(\tau)}{\tau}\,  \dd \tau\right)  -1\right)   {\ln \left\{ 1- z t \left(c_1 + \int^t \frac {b_2(\tau)}{\tau}\, \dd \tau\right)\right\} }   }$\\[10pt]
      $\displaystyle{\qquad  \qquad   \qquad \times \left( c_2 +  \int^t  t  \left(1-z t  \left(c_1 + \int^\tau  \frac{b_2(\tau') }{\tau'}\,\dd \tau'\right)  \right)^{-1}  \dd \tau\right)}$\\[12pt]
          $\displaystyle{\quad  y(t)=x^3(t)\, \frac 1 z \, {\ln \left\{ 1- z t \left(c_1 + \int^t \frac {b_2(\tau)}{\tau}\,  \dd \tau\right)\right\} }  }$
    \\[12pt]

\hline
\\[-6pt]
$\bullet$  {\bf  Case II} \quad  $a(x)=      x^{-1}$\quad arbitrary $b_1(t)$  \quad $ \xii(x)=\ {x^{-1}}$
     \\[3pt]
 $\displaystyle{\quad  \omega=\dfrac{ 1}{x y}\,\dd x\wedge \dd y \qquad  h_{z,1}=\frac{y}{x}\qquad   
h_{z,2}=x\ln x\,\frac{\exp\bigl(-\frac{zy}{x}\bigr)-1}{zy} \qquad h_{z,3}= \frac{x\ln x}{y}\exp\biggl(-\frac{zy}{x}\biggr) }$
   \\[7pt]
 $\displaystyle{\quad   \frac{\dd x}{\dd t}=y +  b_1(t)\,  x \ln x\left(  x \, \frac{\exp\bigl(\frac{zy}{x}\bigr)-1}{zy} -1\right)  \exp\biggl(-\frac{zy}{x}\biggr) - b_2(t)\,\frac{\bigl(x^2+ z x y\bigr)\ln x}{y}  \exp\biggl(-\frac{zy}{x}\biggr)   }$
    \\[12pt]
     $\displaystyle{\quad      \frac{\dd y}{\dd t}=\frac {y^2}x + b_1(t)\left(  x \, \frac{1-\exp\bigl(-\frac{zy}{x}\bigr)}{z} \,(1+\ln x)  - y \exp\biggl(-\frac{zy}{x}\biggr)  \ln x\right)- b_2(t)\bigl(x+(x + zy)\ln x \bigr)  \exp\biggl(-\frac{zy}{x}\biggr)    }$
    \\[12pt]
    $\displaystyle{\quad   \ln x(t)= \frac 1 z \, {\ln \left\{ 1- z \eee^{\gam(t)}  \left(c_1 + \int^t   \eee^{-\gam(\tau)} {b_2(\tau)} \,  \dd \tau\right)\right\} }   \left( \eee^{-\gam(t)} -z \left(c_1 + \int^t   \eee^{-\gam(\tau)} {b_2(\tau)} \,  \dd \tau\right)   \right)       }$\\[10pt]
          $\displaystyle{\qquad  \qquad   \qquad \times    \left\{ c_2 +  \int^t   \left(\eee^{-\gam(\tau)}-z   \left(c_1 + \int^\tau \eee^{-\gam(\tau')}b_2(\tau') \,\dd \tau'\right)  \right)^{-1}  \dd \tau \right\}}$\\[12pt]
         $\displaystyle{\quad   y(t) =x(t)\,  \frac 1 z \, {\ln \left( 1- z \eee^{\gam(t)}  \left(c_1 + \int^t   \eee^{-\gam(\tau)} {b_2(\tau)} \,  \dd \tau\right)\right) }  }$
    \\[12pt]

 \hline
\\[-6pt]
$\bullet$  {\bf  Case III} \quad $a(x)=    \alpha\,  x^{-1}\ (\alpha\ne 1)$\quad arbitrary $b_1(t)$  \quad $ \xii(x)=\ {x^{-\alpha}}$
     \\[3pt]
 $\displaystyle{\quad  \omega=\dfrac{ 1}{x^\alpha y}\,\dd x\wedge \dd y \qquad  h_{z,1}=\frac{y}{x^\aaa} \qquad h_{z,2}=x\, \frac{\exp\bigl(- {zy x^{-\aaa}} \bigr)-1}{zy(1-\aaa)} \qquad h_{z,3}= \frac{x }{y(1-\alpha)}\,\exp\bigl(- {zy x^{-\aaa}} \bigr) }$
   \\[8pt]
 $\displaystyle{\quad   \frac{\dd x}{\dd t}=y +  b_1(t)\,  \frac{x}{1-\aaa}  \left(  x^\aaa \,  \frac{\exp\bigl({zy x^{-\aaa}} \bigr)-1}{zy } -1\right)  \exp\bigl(- {zy x^{-\aaa}} \bigr) - b_2(t)\,\frac{x^{1+\alpha}+ z x y}{(1-\alpha)y}\,  \exp\bigl(- {zy x^{-\aaa}} \bigr) }$
    \\[12pt]
$\displaystyle{\quad   \frac{\dd y}{\dd t}=\frac {\aaa\, y^2}x +  b_1(t)\,  \frac{1}{1-\aaa}  \left(  x^\aaa \,  \frac{\exp\bigl({zy x^{-\aaa}} \bigr)-1}{z } -\aaa y\right)  \exp\bigl(- {zy x^{-\aaa}} \bigr) - b_2(t)\,\frac{x^{\alpha}+ z \alpha y}{1-\alpha }\,  \exp\bigl(- {zy x^{-\aaa}} \bigr)  }$
    \\[12pt]
    $\displaystyle{\quad     x^{1-\alpha}(t)=(1-\alpha) \frac 1 z \, {\ln \left\{ 1- z \eee^{\gam(t)}  \left(c_1 + \int^t   \eee^{-\gam(\tau)} {b_2(\tau)} \,  \dd \tau\right)\right\} }   \left( \eee^{-\gam(t)} -z \left(c_1 + \int^t   \eee^{-\gam(\tau)} {b_2(\tau)} \,  \dd \tau\right)   \right)       }$\\[10pt]
          $\displaystyle{\qquad  \qquad   \qquad \times    \left\{ c_2 +  \int^t   \left(\eee^{-\gam(\tau)}-z   \left(c_1 + \int^\tau \eee^{-\gam(\tau')}b_2(\tau') \,\dd \tau'\right)  \right)^{-1}  \dd \tau \right\}}$\\[12pt]
         $\displaystyle{\quad   y(t) =x^\alpha(t)\,  \frac 1 z \, {\ln \left( 1- z \eee^{\gam(t)}  \left(c_1 + \int^t   \eee^{-\gam(\tau)} {b_2(\tau)} \,  \dd \tau\right)\right) }  }$
    \\[12pt]
     
 \hline

\hline
\end{tabular}
 \end{center}
\label{table2}
}
\end{table}



Propositions~\ref{prop5} and \ref{prop6} state the most general results of this work, covering all the previous systems obtained so far. They  can be applied to particular cases  along the same lines as in Section~\ref{s5} by selecting  precise functions $a(x)$, $b_1(t)$ and $b_2(t)$. The main results concerning the extensions of the systems in Sections~\ref{s51}--\ref{s53} are presented in Table~\ref{table2}, which correspond to the deformation of the $\mathfrak{h}_4$-LH systems shown in Table~\ref{table1}, thus recovered when $z\to 0$. It is clear that  a further analysis of the perturbations at the first-order in $z$ can be performed in a similar way as in Section~\ref{s5}.

\section{Higher-dimensional deformed generalized Buchdahl  equations}
  \label{s8}

As a final stage, it seems pertinent to discuss in more detail the mathematical and physical roles played by the quantum deformation parameter $z$.

In general, given a system,  considering a quantum deformation  by introducing $z$ (or $q=\eee^z$)  implies  dealing with an additional degree of freedom which, in turn, can be regarded as a modification of the initial system. One approach is to interpret this presence as a (integrable) perturbation of the initial system as we have considered here, but, in addition, it also allows the construction of  analytical models from experimental results. In other words, in some cases it would be possible to fix a certain value of $z$ in some quantum algebra that matches with the data for some model, arriving to an underlying quantum group symmetry and, therefore, with analytical expressions. For instance, this was exactly the procedure used~in~\cite{optics} 
 to determine the spectrum in quantum optical models  and  in~\cite{fermion} to describe  fermion-boson interactions in a nuclear physics context.

Furthermore, the introduction of $z$ usually leads to a coupling of the differential equations of the initial system,  as shown  by the explicit expressions in Sections~\ref{s5} and \ref{s7}; this   has consequences with respect to linearization and maximal symmetry which are broken.

 Beyond these comments, a quantum deformation, i.e.,~a Poisson--Hopf deformation of LH systems in our framework, has profound implications when constructing higher-dimensional systems.  In particular, let us address this point by taking the oscillator $\mathfrak{h}_4$-LH algebra of Section~\ref{s6} and its quantum deformation  $\mathfrak{h}_{z,4}$ of Section~\ref{s7} in canonical variables $(q,p)$, as they give rise to the most general systems of this work.
 
 The tool which enables one to obtain  higher-dimensional systems is the coproduct map which for any LH algebra is always (trivial) primitive and denoted by $\Delta$. Let    $(v_1,v_2,v_3,v_0)$ be a basis of  $\mathfrak{h}_4$,  fulfilling the Lie brakets 
\be
[v_2,v_1]= - v_1  ,\qquad [v_2,v_3]=   v_3  ,\qquad [ v_3,v_1]= v_0  ,
\qquad [ v_0,\,\cdot\,]=0 ,
\nonumber
\ee
so formally similar to (\ref{g2}). The Hopf structure is determined by the coproduct $(i=0,1,2,3)$
\be
\Delta(v_i)=v_i\otimes 1 + 1 \otimes v_i  .
\nonumber
\ee
 If we denote the `one-particle' symplectic representation of $\mathfrak{h}_4$ (\ref{g1}) by 
 \be
D( v_i)= h_{i}(q_1,p_1):=  h_{i}^{(1)}  ,
\nonumber
\ee
then the `two-particle' representation is obtained as~\cite{BHLS}
\be
 \begin{split}
(D\otimes D)(\Delta( v_1))&= h_1(q_1,p_1)+h_1(q_2,p_2)= - q_1 - q_2:=  h_{1}^{(2)},\\[2pt]
(D\otimes D)(\Delta( v_2))&= h_2(q_1,p_1)+h_2(q_2,p_2)=q_1p_1+q_2p_2:=  h_{2}^{(2)},\\[2pt]
(D\otimes D)(\Delta( v_3))&= h_3(q_1,p_1)+h_3(q_2,p_2)=  p_1+ p_2:=  h_{3}^{(2)},\\[2pt]
(D\otimes D)(\Delta( v_0))&= h_0(q_1,p_1)+h_0(q_2,p_2)= 1+1:=  h_{0}^{(2)}.
  \end{split}
\nonumber
\ee
These Hamiltonian functions satisfy the same commutation relations (\ref{g2}) with respect to the canonical symplectic form $\omega^{(2)}_{\rm can}=\dd q_1\wedge \dd p_1+ \dd q_2\wedge \dd p_2$. In the same way, one can construct higher-dimensional representations. The relevant point is that any system with  $\mathfrak{h}_4$-LH algebra symmetry in any dimension is equivalent to considering several copies of the initial system, which is a property well-known for Lie systems; in other words, they can trivially be reduced to the initial  `one-particle' one. In our case, the two-particle version of the $t$-dependent Hamiltonian $h_t$ (\ref{g3}) reads as
\be 
 \begin{split}
 h_t^{(2)}&= h^{(2)}_1 + b_1(t)  h^{(2)}_2  + b_2(t)  h^{(2)}_3     = h_t(q_1,p_1)+h_t(q_2,p_2)  \\[2pt]
&=\bigl(  - q_1+  b_1(t)  q_1p_1 +  b_2(t)  p_1\bigr)+\bigl(  - q_2+  b_1(t)  q_2p_2 +  b_2(t)  p_2\bigr).
 \end{split}
\nonumber
\ee
 
This situation  changes drastically when a Poisson--Hopf deformation is introduced, since the deformed coproduct $\Delta_z$ naturally entails a coupling of the representation~\cite{Ballesteros6,BCFHL}. From the deformed coproduct $\Delta_z$ (\ref{h1})  and representation $D_z$ (\ref{h3}) of  $\mathfrak{h}_{z,4}$, we obtain its    `two-particle' representation in the form
\be
 \begin{split}
(D_z\otimes D_z)(\Delta_z( v_1))&= h_{z,1}(q_1,p_1)+h_{z,1}(q_2,p_2)= - q_1 - q_2:=  h_{z,1}^{(2)},\\[2pt]
(D_z\otimes D_z)(\Delta_z( v_2))&= h_{z,2}(q_1,p_1)\eee^{-z h_{z,1}(q_2,p_2)}+h_{z,2}(q_2,p_2) \\[2pt]
&=   \left( \frac{ \eee^{z q_1} -1}{z}\right) p_1 \eee^{z  q_2 }+ \left( \frac{ \eee^{z q_2} -1}{z}\right) p_2:=  h_{z,2}^{(2)},\\[2pt]
(D_z\otimes D_z)(\Delta_z( v_3))&= h_{z,3}(q_1,p_1)\eee^{-z h_{z,1}(q_2,p_2)}+h_{z,3}(q_2,p_2)\\[2pt]
 &\qquad +z h_{z,2}(q_1,p_1)\eee^{-z h_{z,1}(q_2,p_2)} h_{z,0}(q_2,p_2) \\[2pt]
 &=    \eee^{z q_1} p_1 \eee^{z  q_2 } +   \eee^{z q_2} p_2+ z \left( \frac{ \eee^{z q_1} -1}{z}\right) p_1  \eee^{z q_2}  :=  h_{z,3}^{(2)},\\[2pt]
(D_z\otimes D_z)(\Delta_z( v_0))&= h_{z,0}(q_1,p_1)+h_{z,0}(q_2,p_2)= 1+1:=  h_{z,0}^{(2)}.
  \end{split}
\nonumber
\ee
They fulfill the  deformed commutation relations (\ref{h4}) with respect to   $\omega^{(2)}_{\rm can}$. The `two-particle' version of the Hamiltonian  $h_{z,t}$ (\ref{hh66}) turns out to be 
\be
h^{(2)}_{z,t}=h^{(2)}_{z,1} + b_1(t)  h^{(2)}_{z,2}   + b_2(t)  h^{(2)}_{z,3},\nonumber
\ee
and, consequently, is no longer the sum of two copies of (\ref{hh66})  as $h_{z,t}(q_1,p_1)+h_{z,t}(q_2,p_2)$.
The corresponding Hamilton equations are given by
\begin{equation} 
\begin{split}
\frac{\dd q_1}{\dd t}&=  b_1(t) \biggl( \frac{ \eee^{z q_1}-1}{z} \biggr)  \eee^{z q_2 } +b_2(t) \bigl( 2\, \eee^{z q_1}  -1\bigr)\eee^{z q_2}, \\[2pt]
\frac{\dd p_1}{\dd t}&=1 - b_1(t) \eee^{z q_1} \eee^{z q_2} p_1 - 2 z b_2(t)  \eee^{z q_1} \eee^{z q_2} p_1 ,\\[2pt]
\frac{\dd q_2}{\dd t}&=  b_1(t) \biggl( \frac{ \eee^{z q_2}-1}{z} \biggr)   +b_2(t)  \eee^{z q_2}, \\[2pt]
\frac{\dd p_2}{\dd t}&=1 - b_1(t)\eee^{z q_2} \bigl( \bigl(\eee^{z q_1}  -1\bigr) p_1 +p_2 \bigr)-  z b_2(t)  \eee^{z q_2} \bigl( \bigl(2\eee^{z q_1}  -1\bigr) p_1 +p_2 \bigr) ,
\end{split}
\nonumber
\end{equation}
showing that there are coupling terms and the resulting system is intrinsic in the sense that it is completely different from the initial one (\ref{h8}), and  hence finding the corresponding solutions should be seen as a new problem.



\section{Concluding remarks}
\label{concl}

In this work, using the general approach proposed in \cite{Bernoulli}, the generalized Buchdahl equation has been revisited from the perspective of LH systems. Although the Buchdahl equation can be solved directly by means of integrating factors (and is further linearizable by point transformations), its reformulation in terms of a  book $\frak{b}_2$-LH system  is of interest, as the solution method can be extended to Poisson--Hopf deformations from the quantum algebra $\frak{b}_{z,2}$, providing a systematic procedure to determine the general solution of differential equations that, in general, do not admit more than one Lie point symmetry and are not related to exact equations. Further, truncation of the series expansion in the quantum deformation parameter $z$ provides perturbations of the equation of arbitrary order, that under certain circumstances can also be solved explicitly. Certain special cases of the generalized Buchdahl equation and their quantum deformations have been analyzed, as well as the first-order approximations in the deformation parameter $z$.   The next natural step, namely extending the underlying Lie algebra to the oscillator algebra $\mathfrak{h}_4\supset \frak{b}_2$ provides additional generalizations of the Buchdahl equation that still preserve the property of allowing an explicit solution, including the quantum deformations $\mathfrak{h}_{z,4}\supset \frak{b}_{z,2}$. An interesting question in this context is whether equivalence criteria for the first-order approximation of the deformed equations can be obtained, eventually making possible the obtainment of canonical forms of such perturbations and simplifying the computation of exact solutions.

To summarize the results obtained in this paper, the most general cases have been presented in Propositions~\ref{prop5} and \ref{prop6}, by considering the deformed oscillator algebra $\mathfrak{h}_{z,4}$. From them, the different  generalized systems of Buchdahl equations can be recovered through the non-deformed limit $z\to 0$ or by setting the additional $t$-dependent coefficient $b_2(t)=0$, as shown in the following diagram:
\medskip

{\small
\centerline{ {\xymatrix@C=0.5em{&
  {\begin{array}{c}\mbox{Extended generalized} \\ \mbox{Buchdahl equations:}\ \mathfrak{h}_{4} \\ \mbox{Proposition~\ref{prop4}} \\ \mbox{Table~\ref{table1}}  \end{array} } 
    \ar[rd]^{ \displaystyle{ \ b_2(t)=0  }}&&\\
  {\begin{array}{c} \mbox{{ }}\\[-6pt]   \mbox{Deformed extended generalized} \\ \mbox{Buchdahl equations:}\ \mathfrak{h}_{z,4} \\ \mbox{Propositions~\ref{prop5} and \ref{prop6} }   \\ \mbox{Table~\ref{table2}}  \end{array} }
  \ar[rd]^{ \displaystyle{ \  b_2(t)=0}} \ar[ur]^{  \displaystyle{\! \! \! \!\! \! \! \!z\to 0 }}   && {\begin{array}{c}\mbox{Generalized} \\ \mbox{Buchdahl equations:}\ \mathfrak{b}_{2} \\ \mbox{Proposition~\ref{prop1} } \\ \mbox{Sections~\ref{s31}--\ref{s33}}\end{array} } \\
   & {\begin{array}{c}\mbox{{ }}\\[-6pt]\mbox{Deformed generalized} \\ \mbox{Buchdahl equations:}\ \mathfrak{b}_{z,2} \\ \mbox{Propositions~\ref{prop2} and~\ref{prop3}}\\ \mbox{Sections~\ref{s51}--\ref{s53}} \end{array} }  \ar[ru]^{\! \! \! \!\! \! \! \!  \displaystyle{   z\to 0}}&}}}
}
\medskip

As a general remark, it is worthy to be observed that any perturbation at an arbitrary order in $z$ of the generalized Buchdahl equation (in particular, \eqref{e19}) always admits a Lagrangian formulation, as actually happens with any scalar second-order ODE \cite{San}. For example, restricting to the case of perturbations of the type 
\begin{equation}\label{vao}
 \frac{\dd^2 x}{\dd t^2}\simeq\frac 3 x
 \left(\frac{\dd x}{\dd t}\right)^2+\frac 1t \,\frac{\dd x}{\dd t} + \phi_z(t,x)\,\frac{\dd x}{\dd t} 
\end{equation} 
such that $\lim_{z\rightarrow 0}\phi_z(t,x)=0$, setting $\displaystyle u = {\partial \mathcal{L}}/{\partial \dot{x}^2}$, a (nonstandard) Lagrangian $\mathcal{L}$ can be obtained as a solution of the first-order linear partial differential equation 
\begin{equation}\label{var}
\frac{\partial u}{\partial t}+\dot{x}\,\frac{\partial u}{\partial x}+\left(\frac{3}{x}\,\dot{x}^2+\left(\frac{1}{t}+\phi_z(t,x)\right)\dot{x}\right)\frac{\partial u}{\partial \dot{x}}+\left(\frac{6}{x}\,\dot{x}+\frac{1}{t}+\phi_z(t,x)\right)u=0,
\end{equation}
which is deeply connected with the Jacobi multipliers \cite{Nuc08}. Clearly, an admissible solution must satisfy the constraint that $\mathcal{L}_0=\lim_{z\rightarrow 0}\mathcal{L}(z,t,x,\dot{x})$ provides a (nonstandard) Lagrangian for the undeformed Buchdahl equation \eqref{a1}. It can be easily verified that the Lagrangian $\mathcal{L}_0=t^3x^6\dot{x}^{-2}$ given in \cite{Nikiciuk} is a particular solution of \eqref{var} for $\phi_z(t,x)=0$, as well as the alternative Lagrangian $\mathcal{L}_1=(t^3x^6)/(\dot{x}^{2}+kt^2x^6)$, where $k$ is an arbitrary nonzero constant.\footnote{In this context, we observe that the second Lagrangian $L_2=(k \dot{x}^2t^3x^6+t)^{-1}$ given in \cite{Nikiciuk} does not provide the equation \eqref{a1}, but the equation $\frac{\dd^2 x}{\dd t^2}=-\left(\frac 3 x \left(\frac{\dd x}{\dd t}\right)^2+\frac 1t \,\frac{\dd x}{\dd t}\right)$.}

The solution procedure is valid for other types of scalar ordinary differential equations that can be expressed as LH systems based on the book algebra $\mathfrak{b}_2$, encompassing, among others, complex Bernoulli equations with real parameter functions, some type of  Lotka--Volterra systems and various oscillator systems, as well as their corresponding quantum deformation \cite{Bernoulli,ECH}.   The same holds for equations leading to LH systems governed by the oscillator algebra $\mathfrak{h}_4$. In this context, a physically relevant class of differential equations that deserves a deeper analysis in connection with the LH formalism is given by the three-dimensional Hamiltonians associated to metric tensors in  (3+1)-dimensions, in the specific context of integrable cosmological models \cite{Ros}.

Finally, concerning the extension of the method of exact solutions to other LH systems (and their corresponding quantum deformations) based on Lie algebras that contains $\mathfrak{b}_2$ as subalgebra, let us mention that the remaining possibilities in the classification in \cite{LH2015,BHLS} are the simple Lie algebra $\mathfrak{sl}(2,\mathbb{R})$ and the two-photon one $\mathfrak{h}_6\supset \mathfrak{h}_4\supset \mathfrak{b}_2$. However, no exact solution is  yet known to be obtainable for such LH systems, so that the LH approach considered merely provides $t$-independent constants of the motion and superposition rules~\cite{BHLS,Ballesteros6,BCFHL},  from which eventually the generic solution can be derived once a sufficient number of particular solutions has been determined. Detailed analysis of these further generalizations is currently in progress. 


\section*{Acknowledgements}
\phantomsection
\addcontentsline{toc}{section}{Acknowledgments}

{ \small
R.C.S.~and F.J.H.~have been partially supported by Agencia Estatal de Investigaci\'on (Spain) under  the grant PID2023-148373NB-I00 funded by MCIN/AEI/10.13039/501100011033/FEDER, UE.   F.J.H.~acknowledges support  by the  Q-CAYLE Project  funded by the Regional Government of Castilla y Le\'on (Junta de Castilla y Le\'on, Spain) and by the Spanish Ministry of Science and Innovation (MCIN) through the European Union funds NextGenerationEU (PRTR C17.I1).    The authors also acknowledge the contribution of RED2022-134301-T funded by MCIN/AEI/10.13039/501100011033 (Spain).
}

 \newpage

\section*{Appendix A. Symmetry generators of the equation \eqref{b01}}
\label{AA}
\setcounter{equation}{0}
\renewcommand{\theequation}{A.\arabic{equation}}

\phantomsection
\addcontentsline{toc}{section}{Appendix A. Symmetry generators of the equation \eqref{b01}}

  The Lie point symmetries of the generalized Buchdahl equations are   obtained using the standard Lie symmetry method \cite{OLV}. 

  Let us denote $A(x)=\int^x a(\xi) {\rm d}\xi$ and $B(t)=\int^t b(\tau) {\rm d}\tau$. The Lie symmetry generators are given by
\begin{equation*}
\begin{split}
Y_1&= {\rm e}^{A(x)}\frac{\partial}{\partial x},\qquad Y_2= {\rm e}^{A(x)}\left(\int {\rm e}^{-A(x)}{\rm d}x\right)\frac{\partial}{\partial x},\\[2pt]
 Y_3&= {\rm e}^{-B(t)}\frac{\partial}{\partial t}, \qquad Y_4 ={\rm e}^{-B(t)}\left(\int {\rm e}^{B(t)}{\rm d}t\right)\frac{\partial}{\partial t},\\[2pt]
 Y_5 &= {\rm e}^{A(x)}\left(\int {\rm e}^{B(t)}{\rm d}t\right)\frac{\partial}{\partial x},  \qquad Y_6   ={\rm e}^{-B(t)}\left(\int {\rm e}^{-A(x)}{\rm d}x\right)\frac{\partial}{\partial t},\\[2pt]
Y_7&=  {\rm e}^{-B(t)} \left( \int {\rm e}^{-A(x)}{\rm d}x \right)\left(\int {\rm e}^{B(t)}{\rm d}t\right)\frac{\partial}{\partial t}+  {\rm e}^{A(x)} \left(\int {\rm e}^{-A(x)}{\rm d}x\right)^2 \frac{\partial}{\partial x},\\[2pt]
Y_8&=  {\rm e}^{-B(t)}\left(\int {\rm e}^{B(t)}{\rm d}t\right)^2\frac{\partial}{\partial t}+{\rm e}^{A(x)}\left( \int {\rm e}^{-A(x)}{\rm d}x \right)\left(\int {\rm e}^{B(t)}{\rm d}t\right)\frac{\partial}{\partial x}.
\end{split}
\end{equation*}
The corresponding nontrivial commutators are 
\begin{equation*}
\begin{split}
\left[Y_1,Y_2\right]&=Y_1,\qquad \left[Y_1,Y_6\right]=Y_3,\qquad \left[Y_1,Y_7\right]=Y_4+2Y_2,\qquad \left[Y_1,Y_8\right]=Y_5,\\[2pt]
  \left[Y_2,Y_5\right]&=-Y_5, \quad\,  \left[Y_2,Y_6\right]=Y_6,\qquad \left[Y_2,Y_7\right]=Y_7,\qquad \left[Y_3,Y_4\right]=Y_3,\qquad \left[Y_3,Y_5\right]=Y_1,\\[2pt]
 \left[Y_3,Y_7\right]&=Y_6,\qquad \left[Y_3,Y_8\right]=2Y_4+Y_2,\qquad \left[Y_4,Y_5\right]=Y_5,\qquad \left[Y_4,Y_6\right]=-Y_6,\\[2pt]
 \left[Y_4,Y_8\right]&=Y_8,\qquad \left[Y_5,Y_6\right]=Y_4-Y_2,\qquad  \ \,\left[Y_5,Y_7\right]=Y_8,\qquad \left[Y_6,Y_8\right]=Y_7, 
\end{split}
\end{equation*}
which are easily seen to generate a Lie algebra isomorphic to $\mathfrak{sl}(3,\mathbb{R})$, implying that the equation is linearizable via  a point transformation \cite{Mah,OLV}. We further observe that two-dimensional Lie point symmetry algebras of canonical types $L_{2,1}^{\rm I}$, $L_{2,1}^{\rm II}$, $L_{2,2}^{\rm I}$ and $L_{2,2}^{\rm II}$ (see \cite{Mah} for details) also implying linearization are respectively generated by the following vector fields:  
\begin{equation*}
\begin{split}
L_{2,1}^{\rm I}\!: & \quad Y_1,\ Y_3,\quad \left[Y_1,Y_3\right]=0,\quad\ \delta =-{\rm e}^{A(t)}{\rm e}^{-B(t)}\neq 0 ,\\[2pt]
L_{2,1}^{\rm II}\!: & \quad Y_1,\ Y_5,\quad \left[Y_1,Y_5\right]=0,\quad\  \delta = 0,\\
L_{2,2}^{\rm I}\!: & \quad Y_2,\ Y_6,\quad \left[Y_2,Y_6\right]=Y_6,\quad \delta =-{\rm e}^{A(x)}{\rm e}^{-B(t)}\left(\int {\rm e}^{-A(x)}{\rm d}x\right)^2\neq 0 ,\\[2pt]
L_{2,2}^{\rm II}\!: & \quad Y_1,\ Y_2,\quad \left[Y_1,Y_2\right]=Y_1,\quad \delta =0 ,
\end{split}
\end{equation*}
where $\delta =\det \left(\begin{array}[c]{cc}
\xi_1 & \eta_1\\
\xi_2 & \eta_2
\end{array}\right)
$ and $Z_i = \xi_i \frac{\partial}{\partial t}+ \eta_i \frac{\partial}{\partial x}$ are the infinitesimal symmetry generators for $i=1,2$.

 The point symmetries of equations \eqref{e6}, \eqref{e13} and \eqref{e19} are computed similarly. On the other hand, equations of type \eqref{vao} admit point symmetries of the type  
 $$
 \displaystyle Z = \beta_1 \frac{\partial}{\partial t}+\left(\frac{\alpha_1}{ x^3}+\alpha_2x \right) \frac{\partial}{\partial x}
 $$ for perturbation terms of the form 
\begin{equation*}
\phi_z(t,x)= \frac{1}{t}\Phi\!\left(z,\frac{1}{\alpha_2}\,t^{\frac{-4\alpha_2}{\beta_1}}\bigl(\alpha_2x^4+\alpha_1 \bigr)\right).
\end{equation*}


\end{document}